\documentclass[twocolumn]{aastex62}
\usepackage{amsmath}
\usepackage{graphicx}
\usepackage{mathtools}

\usepackage{xcolor}

\newcommand{\OIII}{\mbox{[O III]}}
\newcommand{\NII}{\mbox{[N II]}}
\newcommand{\SII}{\mbox{[S II]}}
\newcommand{\Ha}{H$\alpha$}
\newcommand{\Hb}{H$\beta$}
\newcommand{\OIIIHb}{\OIII/H$\beta$}
\newcommand{\NIIHa}{\NII/H$\alpha$}
\newcommand{\kms}{\mbox{km s$^{-1}$}}
\newcommand{\vout}{$v_{\rm out}$}
\newcommand{\rout}{$R_{\rm out}$}
\newcommand{\Mdotout}{$\dot{M}_{\rm out}$}
\newcommand{\Edotout}{$\dot{E}_{\rm out}$}
\newcommand{\pdotout}{$\dot{p}_{\rm out}$}

\newenvironment{nscenter}
 {\parskip=0pt\par\nopagebreak\centering}
 {\par\noindent\ignorespacesafterend}

\shorttitle{Zooming in on AGN-Driven Outflows at $z\sim$~2.2}
\shortauthors{Rebecca L. Davies et al.}

\begin{document}

\title{From Nuclear to Circumgalactic: Zooming in on AGN-Driven Outflows at $z\sim$~2.2 with SINFONI}

\correspondingauthor{Rebecca L. Davies}
\email{rdavies@mpe.mpg.de}

\author{Rebecca L. Davies}
\affiliation{Max-Planck-Institut f\"ur extraterrestrische Physik,
             Giessenbachstrasse, D-85748 Garching, Germany}

\author{N.M. F\"orster Schreiber}
\affiliation{Max-Planck-Institut f\"ur extraterrestrische Physik,
             Giessenbachstrasse, D-85748 Garching, Germany}
\author{D. Lutz}
\affiliation{Max-Planck-Institut f\"ur extraterrestrische Physik,
             Giessenbachstrasse, D-85748 Garching, Germany}
\author{R. Genzel}
\affiliation{Max-Planck-Institut f\"ur extraterrestrische Physik,
             Giessenbachstrasse, D-85748 Garching, Germany}

\author{S. Belli}
\affiliation{Max-Planck-Institut f\"ur extraterrestrische Physik,
             Giessenbachstrasse, D-85748 Garching, Germany}
\affiliation{Harvard-Smithsonian Center for Astrophysics, 60 Garden Street, Cambridge, MA 02138, USA}

\author{T.T. Shimizu}
\affiliation{Max-Planck-Institut f\"ur extraterrestrische Physik,
             Giessenbachstrasse, D-85748 Garching, Germany}
                          
\author{A. Contursi}
\affiliation{Max-Planck-Institut f\"ur extraterrestrische Physik,
            Giessenbachstrasse, D-85748 Garching, Germany}
\affiliation{IRAM, 300 Rue de la Piscine, 38406 Saint Martin D’Hères, Grenoble, France}

\author{R.I. Davies}
\affiliation{Max-Planck-Institut f\"ur extraterrestrische Physik,
             Giessenbachstrasse, D-85748 Garching, Germany}

\author{R. Herrera-Camus}
\affiliation{Max-Planck-Institut f\"ur extraterrestrische Physik,
             Giessenbachstrasse, D-85748 Garching, Germany}
\affiliation{Departamento de Astronomía, Universidad de Concepción, Barrio Universitario, Concepción, Chile}

\author{M.M. Lee}
\affiliation{Max-Planck-Institut f\"ur extraterrestrische Physik,
             Giessenbachstrasse, D-85748 Garching, Germany}

\author{T. Naab}
\affiliation{Max-Planck-Institut f\"ur Astrophysik, 
	    Karl-Schwarzschildstr. 1, D-85748 Garching, Germany}
             
\author{S.H. Price}
\affiliation{Max-Planck-Institut f\"ur extraterrestrische Physik,
             Giessenbachstrasse, D-85748 Garching, Germany}

\author{A. Renzini}
\affiliation{INAF – Osservatorio Astronomico di Padova, Vicolo dell’Osservatorio 5, I-35122 Padova, Italy}

\author{A. Schruba}
\affiliation{Max-Planck-Institut f\"ur extraterrestrische Physik,
             Giessenbachstrasse, D-85748 Garching, Germany}
             
\author{A. Sternberg}
\affiliation{School of Physics \& Astronomy, Tel Aviv University, Ramat Aviv 69978, Israel}

\author{L.J. Tacconi}
\affiliation{Max-Planck-Institut f\"ur extraterrestrische Physik,
             Giessenbachstrasse, D-85748 Garching, Germany}

\author{H. \"Ubler}
\affiliation{Max-Planck-Institut f\"ur extraterrestrische Physik,
             Giessenbachstrasse, D-85748 Garching, Germany}

\author{E. Wisnioski}
\affiliation{Research School of Astronomy \& Astrophysics, Australian National University, Canberra, ACT 2611, Australia}
\affiliation{ARC Centre of Excellence for All Sky Astrophysics in 3 Dimensions (ASTRO 3D), Australia}

\author{S. Wuyts}
\affiliation{Department of Physics, University of Bath, Claverton Down, Bath, BA2 7AY, UK}

\begin{abstract}
We use deep adaptive optics assisted integral field spectroscopy from SINFONI on the VLT to study the spatially resolved properties of ionized gas outflows driven by active galactic nuclei (AGN) in three galaxies at $z\sim$~2.2 -- K20-ID5, COS4-11337 and J0901+1814. These systems probe AGN feedback from nuclear to circumgalactic scales, and provide unique insights into the different mechanisms by which AGN-driven outflows interact with their host galaxies. K20-ID5 and COS4-11337 are compact star forming galaxies with powerful $\sim$1500~\kms\ AGN-driven outflows that dominate their nuclear \Ha\ emission. The outflows do not appear to have any impact on the instantaneous star formation activity of the host galaxies, but they carry a significant amount of kinetic energy which could heat the halo gas and potentially lead to a reduction in the rate of cold gas accretion onto the galaxies. The outflow from COS4-11337 is propagating directly towards its companion galaxy COS4-11363, at a projected separation of 5.4~kpc. \mbox{COS4-11363} shows signs of shock excitation and recent truncation of star formation activity, which could plausibly have been induced by the outflow from COS4-11337. J0901+1814 is gravitationally lensed, giving us a unique view of a compact \mbox{(R = 470~$\pm$~70~pc)}, relatively low velocity ($\sim$650~\kms) AGN-driven outflow. J0901+1814 has a similar AGN luminosity to COS4-11337, suggesting that the difference in outflow properties is not related to the current AGN luminosity, and may instead reflect a difference in the evolutionary stage of the outflow and/or the coupling efficiency between the AGN ionizing radiation field and the gas in the nuclear regions. 
\end{abstract}


\section{Introduction}
There is growing observational evidence for a direct connection between accretion onto supermassive black holes and the evolution of their host galaxies. The masses of stellar bulges are tightly correlated with the masses of their central supermassive black holes \citep[see e.g. reviews in][]{Alexander12, Kormendy13}. The redshift evolution in the black hole accretion rate density of the universe closely resembles the redshift evolution of the star formation rate (SFR) density \citep[e.g.][]{Madau14, Aird15}. These relationships allude to the presence of some physical mechanism which connects the growth of supermassive black holes on parsec scales with the growth of galaxies on kiloparsec scales.

Cosmological simulations and semi-analytic models of galaxy formation have found that feedback from active galactic nuclei (AGN) can efficiently quench star formation (SF) and account for the low baryon conversion efficiency in high mass systems \citep[e.g.][]{DiMatteo05, Springel05, Bower06, Croton06, Hopkins06, Sijacki07, Hopkins08, Somerville08, Schaye15, Beckmann17, Weinberger17, Pillepich18, Dave19, Nelson19}. AGN activity can impact the host galaxy and the surrounding environment by ionizing or photodissociating the gas, heating the halo gas and reducing the rate of cold accretion onto the galaxy, and/or driving fast outflows that eject gas to large galactocentric distances and (temporarily or permanently) remove the fuel for SF \citep[see][and references therein]{Somerville15}. Other possible mechanisms for quenching SF in high mass galaxies include morphological quenching (stabilization of the gas disk due to the presence of a stellar bulge; e.g. \citealt{Martig09}), virial shock heating \citep[e.g.][]{Birnboim03}, and cosmological starvation (reduced rate of cold gas accretion onto the dark matter halo; e.g. \citealt{Feldmann15}).

AGN-driven outflows are expected to be most prevalent during the peak epoch of SF and black-hole growth, at $z\sim$~1-3 \citep[e.g.][]{Madau14, Aird15}. Studies of mass-selected samples of galaxies have found that AGN-driven ionized gas outflows are ubiquitous in the most massive galaxies at this redshift. \citet{NMFS14} investigated the incidence and properties of outflows in seven massive galaxies with high quality \NII+\Ha\ spectra from the \mbox{SINS/zC-SINF} survey. They found that six out of seven galaxies have AGN-driven nuclear outflows, with a typical velocity full width half maximum (FWHM) of $\sim$~1500~\kms. Four out of six objects observed at adaptive optics resolution ($\sim$~1.5 kpc FWHM) have resolved outflows, with intrinsic diameters of 2-3 kpc. Statistical studies by \citet{Genzel14} ($\sim$~100 galaxies) and \citet{NMFS19} ($\sim$~600 galaxies) confirmed that AGN-driven outflows with velocities of 1000-2000~\kms\ are present in the majority of normal star forming galaxies above the Schechter mass, with the incidence reaching as high as 75\% at \mbox{log(M$_*$/M$_\odot$)~$\gtrsim$~11.2}. Such statistical studies of AGN feedback across the normal galaxy population are crucial for constraining the duty cycle of AGN-driven outflows, and the role of AGN feedback in quenching SF.

There have also been extensive studies of ionized outflows in AGN-selected samples of galaxies at high redshift. \citet{Harrison16} showed that the majority of X-ray AGN at 0.6~$<z<$~1.7 drive outflows with velocities exceeding 600~\kms, and that the incidence of outflows increases with the AGN luminosity. \citet{Leung19} found that 17\% of AGN host galaxies from the MOSDEF survey show evidence for fast \mbox{(400-3500~\kms)}, galaxy-wide outflows. Detailed studies of individual strong outflows in luminous AGN have revealed that high velocity material can often be detected to distances of \mbox{$\sim$5-10~kpc}; well beyond the effective radii of the host galaxies \citep[e.g.][]{CanoDiaz12, Harrison12, Brusa15, Cresci15, Carniani16, Zakamska16, HerreraCamus19}. Powerful AGN-driven outflows therefore have the potential to interact with gas on galaxy scales.

It is well established that fast AGN-driven outflows are prevalent at $z\sim$~2, but the impact they have on the evolution of their host galaxies is strongly debated. Powerful outflows have the potential to exhaust the molecular gas reservoirs of their host galaxies faster than SF, suggesting that they could be an important mechanism for quenching SF \citep[e.g.][]{Maiolino12, Cicone14}. The onset of AGN-driven outflows at high stellar masses coincides with a sharp downturn in the average specific SFRs and molecular gas fractions of galaxies \citep{NMFS19}, and there is a high incidence of AGN activity in the post-starburst region of the UVJ diagram \citep{Belli17b}, providing circumstantial evidence to suggest that AGN-driven outflows may be causally connected with SF quenching. However, direct evidence for this link is so far limited. Some studies have reported evidence for suppression of SF and molecular gas content along the trajectories of outflows from luminous quasars, but the SF outside the outflow region appears unaffected, and in some cases SF can be triggered by shocks at the boundary between the outflow and the surrounding interstellar medium (ISM) \citep[e.g.][]{CanoDiaz12, Brusa15, Cresci15, Carniani16, Carniani17} or can even occur in the outflow itself \citep{Maiolino17}. Outflows driven by moderate luminosity AGN at $z$~$\sim$~2 do not appear to have any significant impact on the instantaneous star formation activity of their host galaxies \citep{Scholtz20}.

\begin{table*}[]
\centering
\caption{Physical properties of the galaxies in our sample, derived as described in Section \ref{sec:sample}.}\label{table:galaxy_properties}
\begin{tabular}{lccccccccc}
\hline Galaxy  & RA & DEC & Redshift	& log(M$_*$/M$_\odot$) & $v_{\rm escape}$  & SFR$_{\rm best}$ & SFR Type & $A_V$ & log$\left(\frac{L_{\rm AGN}}{{\rm erg~s}^{-1}} \right)$ \\
   &   &   &  	& & (\kms) & (M$_\odot$ yr$^{-1}$) & & &  \\ \hline
K20-ID5 & 03:32:31.4 & -27:46:23.2 & 2.224 & 11.2 & 720 & 335 & UV + IR & 1.3 & 45.6 \\ 
COS4-11337  & 10:00:28.70 & +02:17:44.8 & 2.096 & 11.3 & 450 & 395 & UV~+~IR & 0.8 & 46.2 \\
COS4-11363 & 10:00:28.71 & +02:17:45.4 & 2.097 &  11.1 & \ldots & 50 & UV~+~IR & 0.9 & \ldots \\ 
J0901 & 09:01:22.4 & +18:14:32.3 & 2.259 & 11.2 & 780 & 200 & IR & 1.2 & 46.3 \\ \hline
\end{tabular}
\end{table*}

In this paper we use deep (5-20 hours on source) adaptive optics assisted near infrared integral field spectroscopy to characterize the resolved properties of AGN-driven outflows in three massive main sequence galaxies at $z\sim$~2.2, and study how the outflows impact their host galaxies and the surrounding environment. Our galaxy sample and datasets are outlined in Section \ref{sec:sample}. The methods used to measure the outflow parameters are described in Section \ref{sec:methods}. The results for \mbox{K20-ID5}, COS4-11337 and J0901+1814 are presented in Sections \ref{sec:id5}, \ref{sec:cos4_11337} and \ref{sec:j0901}, respectively. In Section \ref{sec:discussion} we discuss our results in the context of galaxy evolution, and we present our conclusions in Section \ref{sec:conc}.

Throughout this work we assume a flat $\Lambda$CDM cosmology with \mbox{H$_{0}$ = 70 \kms\ Mpc$^{-1}$} and \mbox{$\Omega_0$ = 0.3}. All galaxy properties have been derived assuming a \citet{Chabrier03} initial mass function.

\section{Sample and Observations}
\label{sec:sample}
\begin{figure}
\includegraphics[scale = 0.6, clip = True, trim = 15 5 35 30]{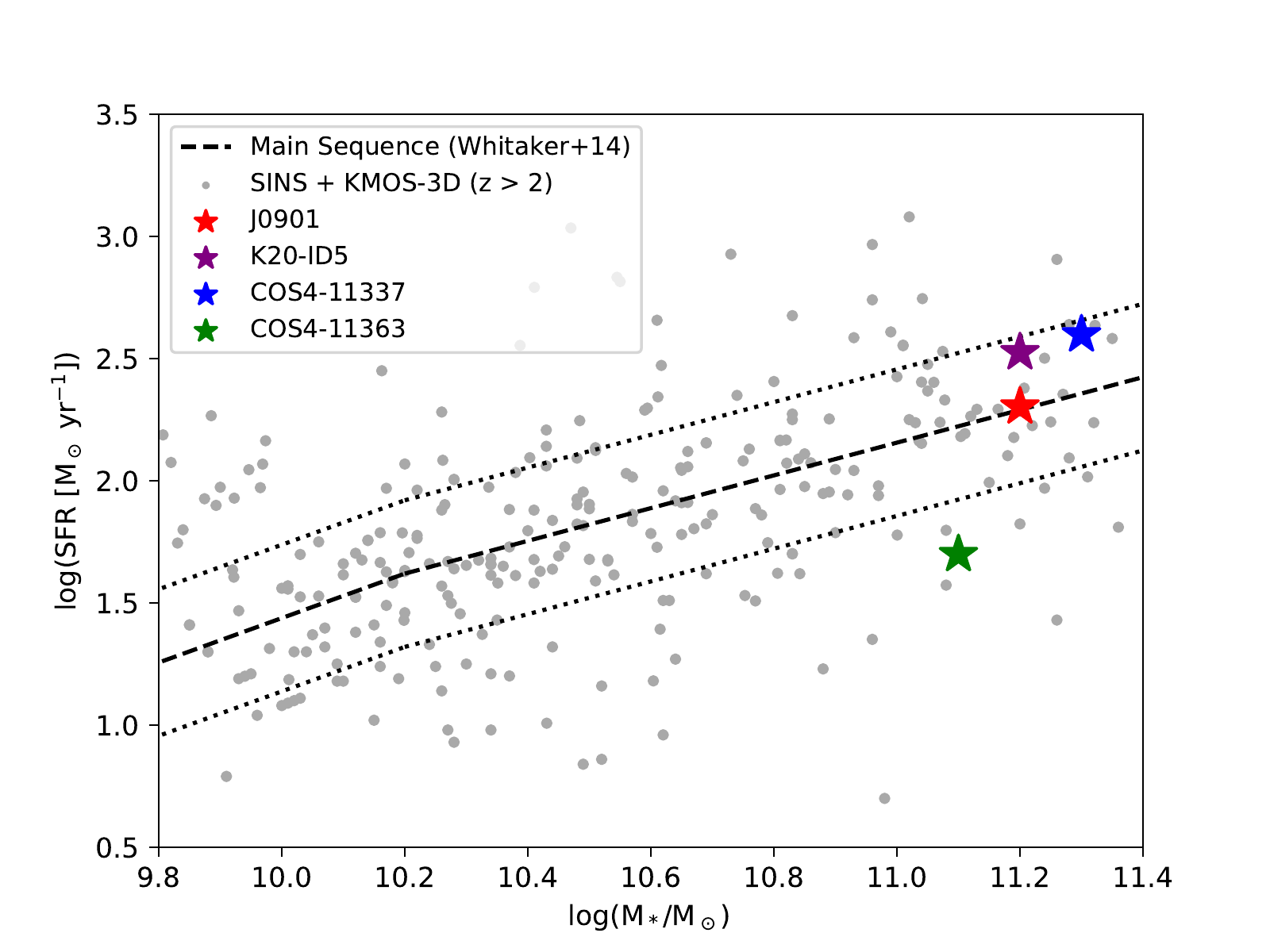} 
\caption{Distribution of the galaxies in our sample (colored stars) in the M$_*$-SFR plane. The black dashed line indicates the SFR main sequence at \mbox{z~=~2-2.5} from \citet{Whitaker14}, and the black dotted lines delineate the $\pm$0.3 dex interval around the main sequence. The grey dots show the distribution of galaxies at z~$\sim$~2-2.6 from the KMOS$^{\rm 3D}$ and SINS/zC-SINF surveys. \label{fig:main_sequence}}
\end{figure}

\mbox{K20-ID5}, \mbox{COS4-11337} and J0901+1814 (J0901 hereafter) were selected from the outflow subsample of the SINS/zC-SINF and KMOS$^{\rm 3D}$ surveys \citep{NMFS14, Genzel14, NMFS19} because they trace AGN feedback on different spatial scales, from nuclear (hundreds of parsecs) to circumgalactic ($>$~5~kpc). The main properties of the galaxies are summarized in Table \ref{table:galaxy_properties}, and Figure \ref{fig:main_sequence} shows where the galaxies are located in the M$_*$-SFR plane, relative to the \mbox{$z$~=~2-2.5} SFR main sequence from \citet{Whitaker14}. The galaxy COS4-11363 is included because it is in a close pair with COS4-11337 and is discussed in detail in Section \ref{sec:cos4_11337}.

Our analysis is primarily based on high spatial resolution rest-frame optical integral field spectroscopy. Each of our galaxies was targeted with the Spectrograph for INtegral Field Observations in the Near Infrared (SINFONI; \citealt{Eisenhauer03, Bonnet04}) on the Very Large Telescope (VLT). The observations were performed using the K band filter \mbox{(1.95-2.45$\mu$m)}, providing integral field datacubes covering the \NII$\lambda$6548, \Ha$\lambda$6563, \NII$\lambda$6584, and \SII$\lambda \lambda$6716, 6731 lines. We used the Laser Guide Star (LGS) Adaptive Optics (AO) mode \citep{Bonnet03}, achieving spatial resolutions of 0.18-0.24'' (FWHM), or 1.5-2~kpc at $z\sim$~2.2. The high spatial resolution is required to measure the spatial extent of the outflows and to separate different line-emitting structures within the galaxies. The reduction of the SINFONI-AO data is described in detail in \citet{NMFS18}. The final K band SINFONI-AO cubes have a pixel scale of 0.05'' and a spectral resolution of $\sim$~85~\kms. 

\begin{table*}
\caption{Summary of the observations used in this paper.}\label{table:obs_summary}
\begin{tabular}{lccccccc}
\hline Galaxy  	& Instrument   & Filter      & t$_{int}$ & Pixel Scale & PSF FWHM & R & Program ID(s) \\ \hline
K20-ID5   & SINFONI + AO  & K    	& 13h40m  & 0.05''       & 0.18'' 	& 3530 & 093.A-0110(B), 097.B-0065(B) \\ 
\ldots	& KMOS & K    	        & 17h30m           & 0.2''      & 0.36'' 	& 3500 & 092.A-0091(A), 095.A-0047(B) \\ 
\ldots 	& SINFONI & H         & 2h00m            & 0.125''      & 0.68'' & 2500 & 074.A-9011 \\  
COS4-11337/11363 & SINFONI + AO & K   & 5h00m            & 0.05''       & 0.24''	 & 3530 & 097.B-0065(A) \\
\ldots	 	& KMOS & K    & 16h35m  & 0.2''     & 0.40''  & 3590 & 093.A-0079(A) \\ 
\ldots	   	& KMOS   & H       & 4h25m   & 0.2''  & 0.66''  & 3470 & 0101.A-0022(A) \\ 
J0901      & SINFONI + AO  & K & 9h30m  & 0.05"    & 0.20'' & 3530 & 093.A-0110(A), 094.A-0568(A)\\
\ldots	& SINFONI & K & 9h00m & 0.125'' & $\sim$0.6'' & 3530 & 092.A-0082(A) \\ 
\ldots	 & LBT/LUCI + AO & H & 2h15m  & 0.118''  & 0.37'' & 5470 & LBT-2018A-C0208-3 \\ \hline
\end{tabular}
\tablecomments{The KMOS data were obtained as part of the KMOS$^{\rm 3D}$ Survey \citep{Wisnioski15, Wisnioski19}. The SINFONI seeing limited and adaptive optics datasets were reduced as described in \citet{NMFS09} and \citet{NMFS18}, respectively. The LUCI longslit data for J0901 are described in Section \ref{subsec:j0901_properties}. The PSF FWHM values were measured from 2D Moffat (KMOS data) or Gaussian (SINFONI and LUCI data) fits to images of standard stars obtained simultaneously (KMOS) or close in time (SINFONI and LUCI) to the science data, with the exception of the J0901 seeing limited SINFONI dataset for which the PSF FWHM was estimated by comparing the R band seeing (0.96'') to the distributions of R band seeing and K band PSF FWHM values from the KMOS$^{\rm 3D}$ survey (Figure 2 in \citealt{Wisnioski19}).}
\end{table*}

Our high resolution K band data are supplemented with seeing limited K band data, and H band data covering the \OIII$\lambda$4959, \OIII$\lambda$5007 and \Hb\ lines. We use the seeing limited K band observations to probe faint, extended \NII+\Ha\ emission that is undetected in the adaptive optics observations. The H band observations assist in decomposing the emission line profiles into multiple kinematic components, because the \OIII$\lambda$5007 line is strong in AGN host galaxies and is not blended even when a broad outflow component is present, unlike \Ha\ which can be strongly blended with \NII. The \Hb\ line, when robustly detected, is used to measure the Balmer decrement. The full set of observations used in this paper is summarized in Table \ref{table:obs_summary}. 

\subsection{K20-ID5}\label{subsec:id5_properties}
K20-ID5 (also known as 3D-HST GS4-30274, \mbox{GS3-19791} and GMASS 0953) is a well studied star-forming galaxy at $z\sim$~2.224. It was identified as a high redshift candidate ($z_{\rm phot}$~$>$~1.7) in the K20 survey of infrared bright galaxies \citep{Cimatti02}, spectroscopically confirmed by \citet{Daddi04}, and followed up with deeper long slit spectroscopy as part of the GMASS survey \citep{Kurk13}, and integral field spectroscopy as part of the SINS/zC-SINF \citep{NMFS09}, KMOS$^{\rm 3D}$ \citep{Wisnioski15} and KASHz \citep{Harrison16} surveys.

\begin{figure*}
\centering
\includegraphics[scale=0.6,clip = True, trim = 30 0 0 0]{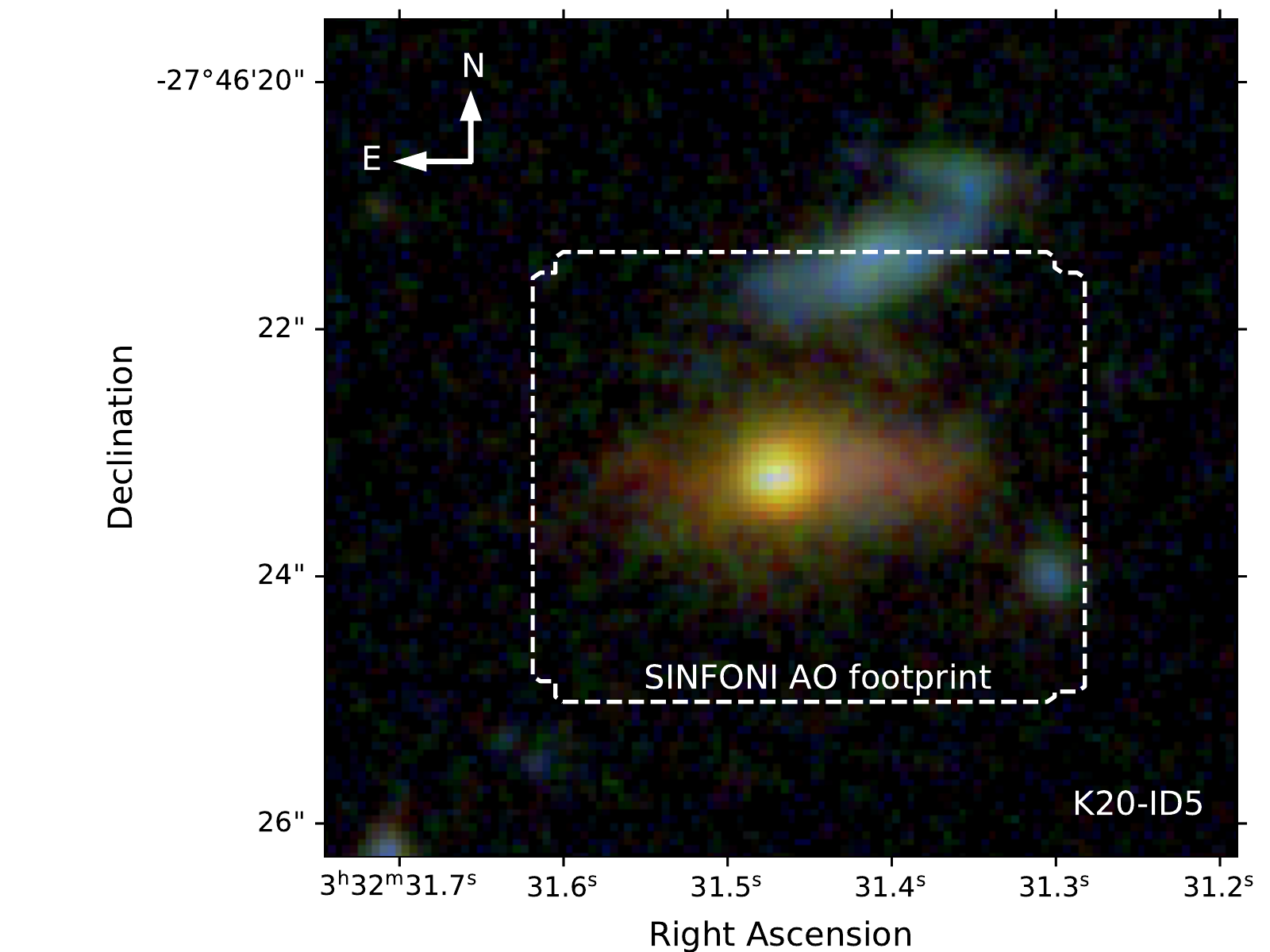} \includegraphics[scale=0.8, clip = True, trim = 0 75 35 110]{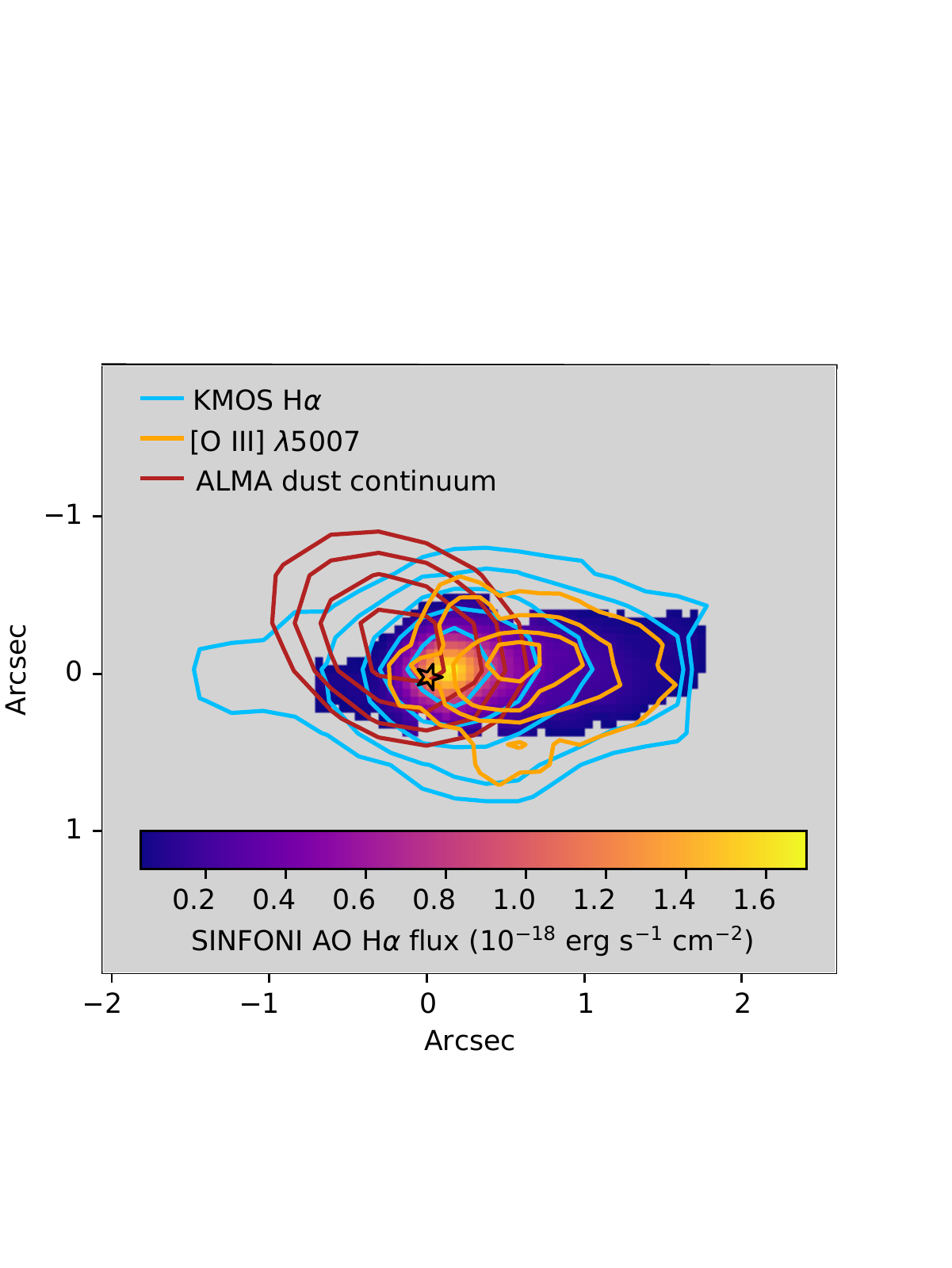}
\caption{Left: K20-ID5 color composite using HST F160W (red) + F125W (green) + HST F814W (blue), with the footprint of the SINFONI-AO data overplotted. The blue galaxy to the north of K20-ID5 is a lower redshift (foreground) system. Right: SINFONI-AO \Ha\ flux map of K20-ID5, with contours of the KMOS \Ha\ emission (blue), \OIII$\lambda$5007 emission (from seeing limited SINFONI data, yellow) and ALMA band 4 dust continuum (brown) overplotted. The black star indicates the kinematic center of the galaxy. \label{fig:id5_footprints}}
\end{figure*}

The left hand panel of Figure \ref{fig:id5_footprints} shows an IJH color HST composite image of the region around K20-ID5. The galaxy has a bright and compact nucleus surrounded by fainter emission which is stronger and more extended on the western side of the galaxy than the eastern side. The blue galaxy to the north of K20-ID5 is a lower redshift foreground object. 

K20-ID5 has a stellar mass of \mbox{log(M$_*$/M$_\odot$) = 11.2} and a UV~+~IR SFR of \mbox{335~M$_\odot$ yr$^{-1}$} \citep{Wuyts11b,Wuyts11}, placing it on the upper envelope of the main sequence (purple star in Figure \ref{fig:main_sequence})\footnote{\citet{Talia18} and \citet{Scholtz20} reported similar values for the stellar mass and SFR (consistent within 0.2 dex), based on SED modelling covering rest frame wavelengths of \mbox{0.1-1000~$\mu$m}.}. The galaxy is classified as an AGN host based on the X-ray luminosity, radio luminosity, mid-infrared colors and optical line ratios \citep[e.g.][]{vanDokkum05, Genzel14, Newman14}. The black hole accretion is driving a galactic wind which has been detected as a broad blueshifted component in the rest-frame optical emission lines \citep{Genzel14, NMFS14, Loiacono19, Scholtz20}, the rest-frame UV absorption lines \citep{Cimatti13}, and tentatively in the CO(6-5) emission line \citep{Talia18}. 

K20-ID5 has an effective radius of \mbox{r$_e$ = 2.5 kpc} \citep{vanderWel14} and is classified as a compact star forming galaxy \citep{vanDokkum15, Wisnioski18}. Molecular gas observations suggest that K20-ID5 has a starburst-like molecular gas depletion time and may therefore quench on a few hundred Myr timescale \citep{Popping17, Talia18, Loiacono19}.

Our SINFONI-AO observations of K20-ID5 cover the white dashed region in the left hand panel of Figure \ref{fig:id5_footprints}. Over the same region we also have deep seeing-limited KMOS K band observations and seeing limited SINFONI H band observations. The right hand panel of Figure \ref{fig:id5_footprints} shows the distribution of the SINFONI-AO \Ha\ flux (mapped in colors), the KMOS \Ha\ and SINFONI \OIII\ flux (measured from single Gaussian fits; blue and yellow contours, respectively), and the dust distribution traced by the ALMA band 4 continuum emission (program 2015.1.00228.S, PI: G. Popping) (brown contours). The SINFONI-AO \Ha\ flux map has a similar asymmetric distribution to the broadband flux, but the KMOS$^{\rm 3D}$ data indicate that \Ha\ emission is indeed present on the fainter eastern side of the galaxy, and further analysis reveals that the ionized gas follows a regular rotation curve on both sides of the nucleus \citep{Wisnioski18}. The ALMA data suggest that there is 2.5~$\times$~10$^8$~M$_\odot$ of dust centered just to the east of the nucleus \citep{Popping17, Talia18}, and the Balmer decrement indicates significant attenuation of nebular emission in the nuclear region of the galaxy (A$_V$~=~2.7; see Section \ref{subsec:id5_energetics}). The large amount of dust could explain the asymmetry in the UV/optical emission, as well as the significant offset between the \Ha\ and \OIII\ flux peaks (also seen in \citealt{Loiacono19} and \citealt{Scholtz20}). 

The obscuration also complicates the measurement of the AGN luminosity. Using the X-ray hardness ratio as a tracer of the obscuration yields a column density of \mbox{log(N$_{\rm H}$/cm$^{-2}$) = 23.2}, corresponding to an intrinsic hard (2-10 keV) X-ray luminosity of \mbox{log(L$_X$/erg s$^{-1}$) = 43.0} and a bolometric luminosity of \mbox{log(L$_{\rm AGN}$/erg s$^{-1}$) = 44.2} (using the bolometric correction adopted by \citealt{Rosario12}; their Equation 1). However, ongoing X-ray spectral modelling efforts suggest that the AGN may be Compton thick (Dalla Mura et al. in prep), indicating that the hard X-ray luminosity is likely to be underestimated. Therefore, we also estimate the AGN luminosity from the \NII\ line (as described in \citealt{NMFS19}), yielding \mbox{log(L$_{\rm AGN}$/erg s$^{-1}$) = 45.9}. This value is much closer to what one would derive from the X-ray flux if the AGN was assumed to be Compton thick. Finally, we use the Spitzer/MIPS 24$\mu$m (rest-frame 7.4$\mu$m) flux density to derive an upper limit on the AGN luminosity. \citet{Lutz04} found a linear correlation between $\nu F_{\nu}$(6$\mu$m) and \mbox{$F$(2-10 keV)}, with a typical slope of 0.23 for Seyfert 2 AGN. The observed mid-infrared flux includes contributions from both SF and AGN activity, and therefore the total $\nu F_{\nu}$ provides an upper limit on the AGN luminosity for K20-ID5. We measure \mbox{log($\nu L_{\nu}$[7.4$\mu$m, AGN]/erg s$^{-1}$) $<$ 45.2}, corresponding to a hard X-ray luminosity of \mbox{log(L$_X$/erg s$^{-1}$) $<$ 44.6} and a bolometric luminosity of \mbox{log(L$_{\rm AGN}$/erg s$^{-1}$) $<$ 46.4}. The upper limit is 0.5 dex higher than the luminosity calculated from \NII. Therefore, we adopt the average of the X-ray and \NII\ estimates as our fiducial AGN luminosity; i.e. \mbox{log(L$_{\rm AGN}$) = 45.6}. 

\subsection{COS4-11337/COS4-11363}
\begin{figure*}
\centering
\includegraphics[scale=0.6,clip = True, trim = 70 10 20 0]{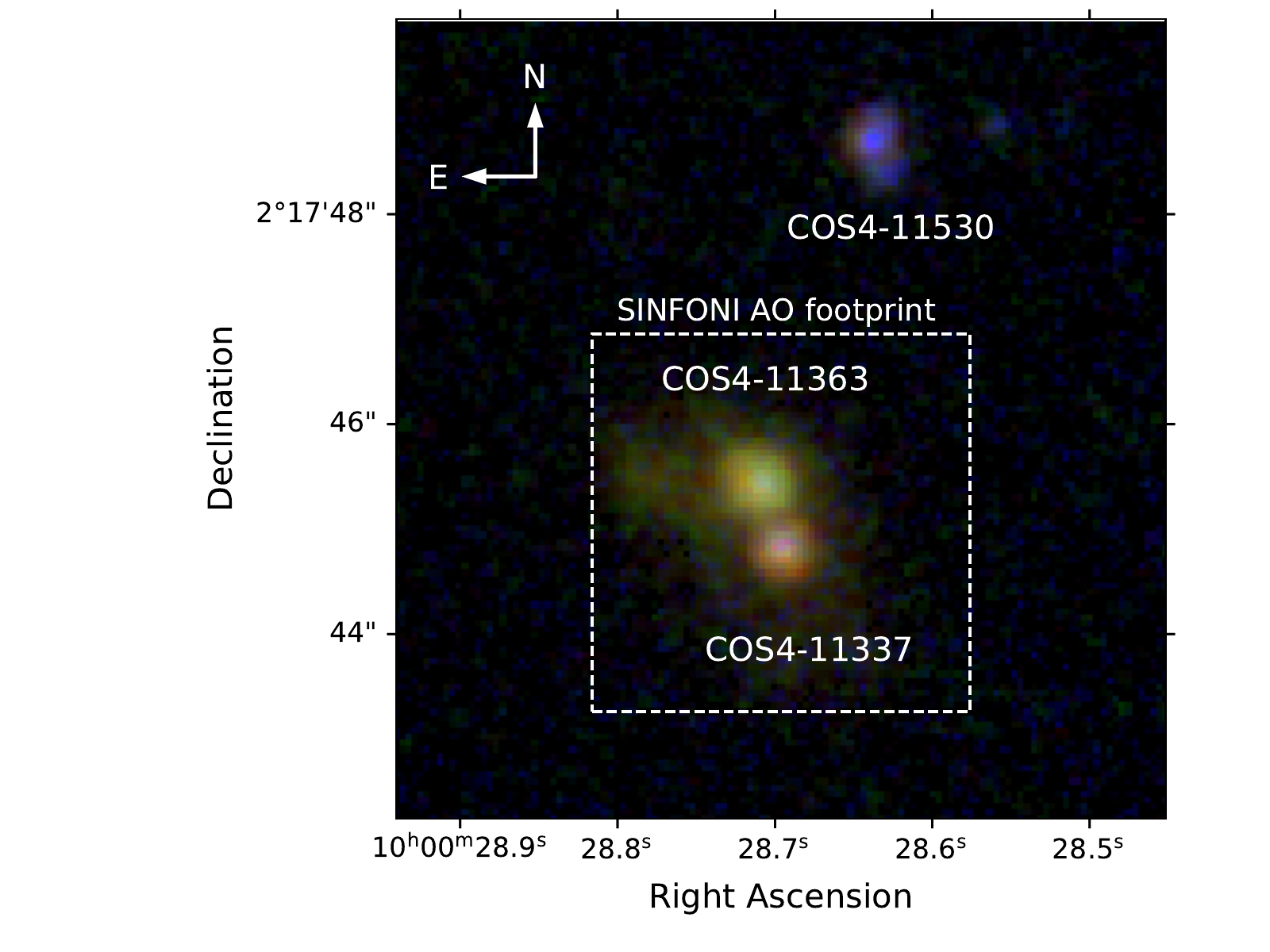} 
\includegraphics[scale=1.1,clip = True, trim = 210 160 30 0]{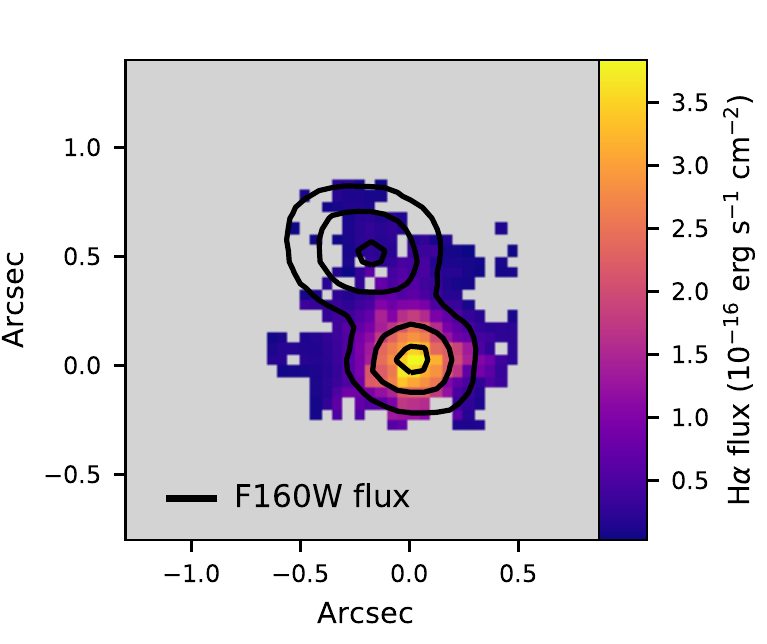}
\caption{Left: Color composite of the \mbox{COS4-11337/COS4-11363/COS4-11530} triplet using HST F160W (red) + F125W (green) + F814W (blue). The white box indicates the region covered by our SINFONI-AO and KMOS observations. Right: Map of the \Ha\ flux from the SINFONI-AO cube, with contours of the F160W emission overlaid (at levels of 10\%, 25\% and 75\% of the peak flux). \label{fig:cos4_11363_footprints}}
\end{figure*}

\begin{sloppypar}
COS4-11337 and COS4-11363 are two massive \mbox{(log(M$_*$/M$_\odot$)~$\sim$~11)}, compact (R$_e \leq$~2.5~kpc; \citealt{vanderWel14}) galaxies which lie very close in redshift ($\Delta v$~=~150~\kms) and have a projected separation of only 0.65'' (5.4 kpc), and may therefore be in the early stages of a major merger. The dwarf starburst galaxy COS4-11530 lies 33 kpc to the north-west of the pair at a very similar spectroscopic redshift ($z$~=~2.097). The left hand panel of Figure \ref{fig:cos4_11363_footprints} shows an IJH color HST composite image of the triplet.
\end{sloppypar}

The stellar masses and $A_V$ values for COS4-11337 and COS4-11363 were initially derived using standard SED modelling (following \citealt{Wuyts11b}), assuming an exponentially declining star formation history and considering e-folding times in the range \mbox{log($\tau$/yr)~=~8.5-10.0}. The best fit galaxy parameters for \mbox{COS4-11337} are \mbox{log(M$_*$/M$_\odot$)~=~11.3} and \mbox{$A_V$~=~0.8}. We found that the fit for COS4-11363 could be significantly improved by adopting either a shorter e-folding time (100~Myr) or a truncated star formation history. Both models give best fit galaxy parameters of \mbox{log(M$_*$/M$_\odot$)~=~11.1} and \mbox{$A_V$~=~0.9}. 

We calculate the IR SFR of the system using the Herschel PEP 160$\mu$m flux which we convert to L$_{\rm IR}$ using the \citet{Wuyts08} SED template. The galaxies are strongly blended in the far infrared imaging, so it is only possible to calculate the total (combined) SFR of the system, which is 424~M$_\odot$~yr$^{-1}$. The SED fitting suggests that COS4-11337 has a $\sim$~10$\times$ higher SFR than \mbox{COS4-11363}, and the \Ha\ flux ratio between the two nuclei is 47, or 14 when removing the contribution of the outflow to the \Ha\ flux of COS4-11337 (see Section \ref{subsec:cos4_11337_outflow_kinem}). Therefore, we divide the IR SFR between the galaxies in a 10:1 ratio; i.e. \mbox{SFR$_{\rm IR, 11337}$ = 385 M$_\odot$~yr$^{-1}$} and \mbox{SFR$_{\rm IR, 11363}$ = 39 M$_\odot$~yr$^{-1}$}. We add the UV SFRs measured from the SED fitting for the individual systems, and obtain UV~+~IR SFRs of 395~M$_\odot$~yr$^{-1}$ and 50~M$_\odot$~yr$^{-1}$ for COS4-11337 and COS4-11363, respectively. 

Based on our derived parameters, \mbox{COS4-11337} lies on the upper envelope of the star formation main sequence (blue star in Figure \ref{fig:main_sequence}), and \mbox{COS4-11363} lies about a factor of three below the main sequence (green star in Figure \ref{fig:main_sequence}).

Grism spectroscopy from the 3D-HST survey \citep{Brammer12, Momcheva16} indicates that \mbox{COS4-11337} has strong \OIII\ emission, while \mbox{COS4-11363} has very weak line emission. \mbox{COS4-11337} is classified as an AGN based on both the optical line ratios \citep{Genzel14} and the hard X-ray luminosity (\mbox{log(L$_X$/erg s$^{-1}$) = 44.5}; \citealt{Luo17}), which corresponds to an AGN bolometric luminosity of \mbox{log($L_{\rm bol, AGN}$/[erg s$^{-1}$]) = 46.2} \citep{Rosario12}. We note that a very similar bolometric luminosity is obtained from the extinction-corrected \OIII\ luminosity, assuming a bolometric correction factor of 600 \citep{Netzer09}. 

Seeing limited H and K band observations of the COS4-11337/COS4-11363 system were obtained as part of the KMOS$^{\rm 3D}$ survey \citep{Wisnioski15, Wisnioski19}. Broad forbidden lines reveal a strong outflow originating from COS4-11337. The beam convolved line emission from COS4-11337 is dominant everywhere line emission is detected, but the \NIIHa\ ratio increases towards the nucleus of \mbox{COS4-11363}, indicating that COS4-11363 does have some (weak) line emission. We targeted the pair with SINFONI~+~AO to allow us to robustly separate the line emission from the two galaxies. The right hand panel of Figure \ref{fig:cos4_11363_footprints} shows a map of the \Ha\ flux from our SINFONI-AO data, with contours of the HST F160W emission overlaid. The \Ha\ flux primarily originates from \mbox{COS4-11337}, but is also detected at the location of \mbox{COS4-11363}, consistent with the results from KMOS$^{\rm 3D}$ and the 3D-HST grism spectroscopy.

\begin{figure*}
\centering
\includegraphics[scale=0.6,clip = True, trim = 25 0 20 0]{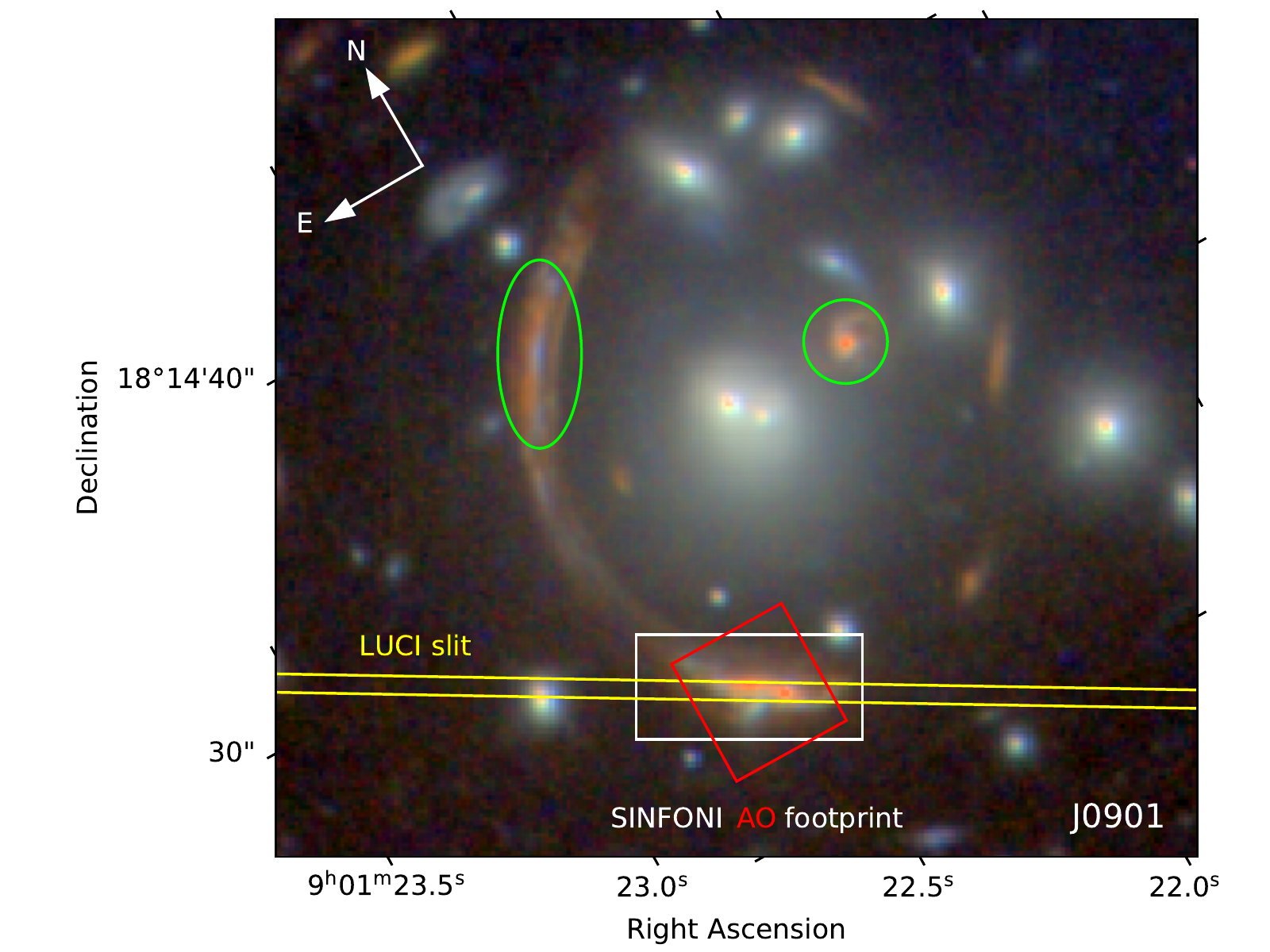}  \includegraphics[scale=0.62, clip = True, trim = 10 0 40 50]{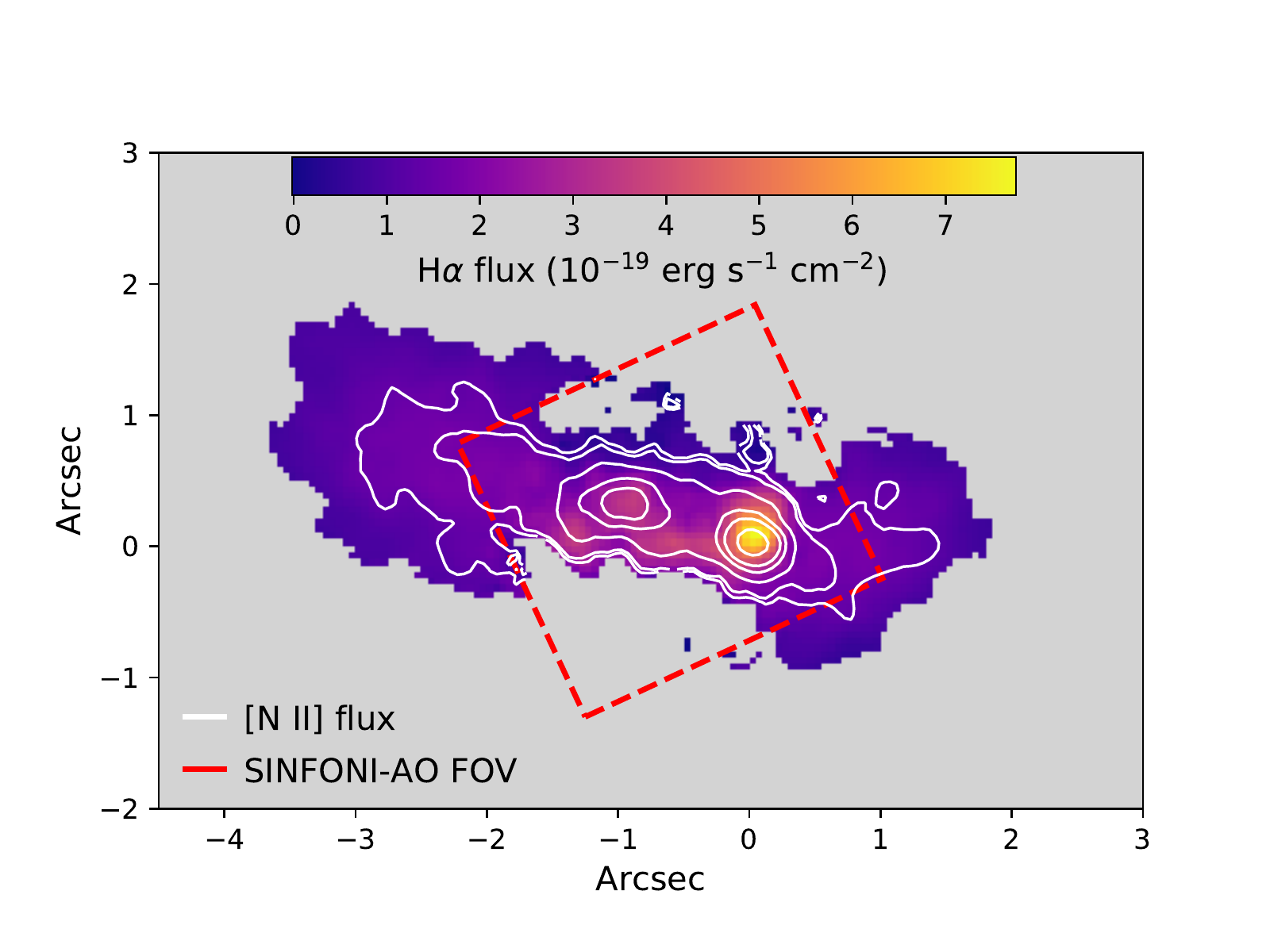} 
\caption{Left: J0901 color composite using HST F160W (red) + F110W (green) + F814W (blue), with the footprints of the LUCI slit (yellow), and the SINFONI seeing limited (white) and AO (red) data overplotted. J0901 is triply imaged by the lensing cluster. The south eastern image is at the center of the SINFONI footprint, and the north eastern and western images are indicated by the green ellipses. Right: Image plane \Ha\ flux map of J0901, constructed by combining the AO data (inside the red rectangle) with the seeing limited data (outer region). The white contours trace the \NII\ flux distribution. \label{fig:j0901_footprints}}
\end{figure*}

\subsection{J0901}\label{subsec:j0901_properties}
SDSS J090122.37+181432.3 (abbreviated to J0901) is a strongly lensed, triply imaged galaxy at $z$~=~2.259, first reported by \citet{Diehl09}. The left hand panel of Figure \ref{fig:j0901_footprints} shows an IJH color HST composite image of the region surrounding the lensing cluster, which is at a redshift of $z$~=~0.346. The green circles indicate the locations of the north eastern (NE) and western (W) images of J0901. The south eastern (SE) image is at the center of the white and red rectangles, which indicate the coverage of our SINFONI seeing limited and adaptive optics datasets, respectively. We targeted the SE arc because the similarly bright NE fold arc is strongly distorted and images only part of the source.

Rest-frame UV and optical spectra of J0901 reveal high \NIIHa\ and \OIIIHb\ ratios \citep{Hainline09} and clear [N~V] emission \citep{Diehl09}, indicative of AGN activity. The \NII+\Ha\ emission line complex shows a clear broad component which likely traces an AGN-driven outflow \citep{Genzel14}.

We obtained K band SINFONI observations of J0901, using the seeing limited mode to probe the extended flux of the galaxy and the adaptive optics mode to obtain a high resolution view of the center of the galaxy. The right hand panel of Figure \ref{fig:j0901_footprints} shows an image plane \Ha\ flux map constructed from the combination of the seeing limited and AO datasets. The fluxes for the pixels inside the red rectangle were measured from the AO data, and the fluxes for the pixels in the outer region were measured from the seeing limited data. The white contours show the spatial distribution of the \NII\ emission.

We also observed J0901 in the H band using the LUCI long slit spectrograph on the Large Binocular Telescope (LBT), aided by the Advanced Rayleigh guided Ground-layer adaptive Optics System \citep[ARGOS;][]{Rabien19}. The observations consist of one 45 minute block from LUCI-1 on \mbox{2018-03-02} and one 45 minute block from each of LUCI-1 and LUCI-2 on \mbox{2018-03-03}. The three blocks were reduced independently using the Flame pipeline \citep{Belli18}, and then combined, weighted by the uncertainty on each pixel. The spatial resolution of the combined dataset is 0.37'', and the spectral resolution calculated from a stack of skylines close to the wavelength of \OIII$\lambda$5007 is 55~\kms. The final 2D spectrum was flux calibrated using the slit alignment star.

The strong gravitational lensing causes magnification and distortion of the light which must be accounted for in order to derive the intrinsic properties of J0901. We use \textsc{Lenstool} and archival HST imaging to build a model for the mass distribution of the lensing cluster, as described in Appendix \ref{appendix:lens_modelling}. This model allows us to convert from observed (image plane) flux distributions to intrinsic (source plane) flux distributions, and derive the intrinsic properties of the galaxy.

\begin{sloppypar}
We measure an intrinsic \OIII\ luminosity of \mbox{log($L_{\rm [O III]}$/[erg s$^{-1}$]) = 43.5}, corresponding to an AGN bolometric luminosity of \mbox{log($L_{\rm bol, AGN}$/[erg s$^{-1}$]) = 46.3} (see Appendix \ref{appendix:j0901_lagn}). We note that the AGN luminosity is relatively uncertain due to both the lack of hard X-ray observations and the uncertainties associated with the source plane reconstruction. We fit SED models to the magnification corrected F160W, F814W and F435W fluxes, yielding a stellar mass of \mbox{log(M$_*$/M$_\odot$) = 11.2} and an $A_V$ of 1.2 (Appendix \ref{appendix:j0901_mstar}). Using the 160$\mu$m flux measurement presented in \citet{Saintonge13} we derive a SFR of 200~M$_\odot$~yr$^{-1}$ (Appendix \ref{appendix:j0901_sfr}) which places J0901 on the SFR main sequence (red star in Figure \ref{fig:main_sequence}).
\end{sloppypar}

\section{Measuring Outflow Parameters}\label{sec:methods}
\subsection{Isolating Emission Associated with Outflows}\label{subsec:separating_outflow_emission}
Ionized gas outflows are generally observed as broadened (FWHM $\gtrsim$ a few hundred km/s) emission line components underneath the narrower emission produced by gas in the disk of the galaxy. The outflow and disk components can be robustly separated for individual spaxels in integral field observations of local galaxies, but even with our deep observations the emission line signal-to-noise (S/N) is only high enough to permit a robust disk-outflow decomposition in the brightest spaxels. Therefore, we calculate the properties of the AGN-driven outflows in K20-ID5, COS4-11337 and J0901 using spectra integrated across the nuclear regions of the galaxies. 

The method used to extract the nuclear spectra is described in detail in Section 2.5.1 of \citet{NMFS19}. In short, each datacube was median subtracted to remove continuum emission, $\sigma$-clipped to remove skyline residuals, and smoothed spatially using a Gaussian filter with a FWHM of 3-4 pixels. For each spaxel in each datacube, we simultaneously fit the \NII$\lambda$6548, \Ha\ and \NII$\lambda$6583 lines as single Gaussians with a common velocity offset and velocity dispersion, and then shifted the spectrum so that the narrow line cores were centered at zero velocity. This velocity shifting removes (and therefore prevents artificial line profile broadening associated with) large scale gravitationally driven velocity gradients, but has minimal impact on the shapes of the outflow line profiles because their line widths are $\sim$~5-10$\times$ larger than the maximum velocity shifts. From the velocity shifted cubes we extracted spectra integrated over the region where broad outflow emission was detected ($\sim$~1-3 kiloparsecs in radius). 

We fit the line profiles in each nuclear spectrum as a superposition of two kinematic components - one for the galaxy and one for the outflow. An example fit (to the nuclear spectrum of COS4-11337) can be seen in the left hand panel of Figure \ref{fig:cos4_11337_nuc_spec_fit}. We again assumed a common velocity offset and velocity dispersion for all lines in each kinematic component, and fixed the \NII$\lambda$6583/\NII$\lambda$6548 and \OIII$\lambda$5007/\OIII$\lambda$4959 ratios to 3 (the theoretical value set by quantum mechanics). 

\subsection{Outflow Extent}\label{subsec:rout}
We use the two component fits to the nuclear spectra to determine which spectral channels are dominated by the outflow component, and integrate the flux over the outflow channels in each spaxel to create maps of the outflow emission. We compare the curves of growth of the outflow emission and the point spread function (PSF) to confirm that the outflows are resolved. We calculate the intrinsic (PSF-corrected) size of the outflows by modelling the observed outflow emission as a 2D Gaussian (representing the intrinsic outflow emission) convolved with the PSF, and adopt the half width at half maximum (HWHM) of the Gaussian as the radius of the outflow. In this modelling we use the empirically-derived average AO PSF from the \mbox{SINS/zC-SINF} AO Survey \citep{NMFS18}, which has sufficiently high S/N that both the core (AO corrected) and wing (uncorrected) components of the PSF can be robustly characterized. The average PSF is constructed from datasets obtained under similar conditions to our targets, and the FWHM measured from the curve of growth of the average PSF is 0.18'', which is similar to the FWHM values measured for our datasets (see Table \ref{table:obs_summary}).

\subsection{Outflow Velocity}\label{subsec:vout}
The outflow velocity is calculated by taking the full width at zero power (FWZP) of the entire \NII+\Ha\ complex, subtracing the velocity separation of the \NII\ doublet lines (1600 \kms), and dividing by two \citep[following][]{NMFS19}. This measurement yields the maximum line-of-sight velocity of the outflowing material, which should be close to the true outflow velocity for a constant velocity wide angle outflow.

\subsection{Mass Outflow Rate (\Mdotout) and Mass Loading Factor ($\eta$)}\label{subsec:mdot}
\begin{sloppypar}
The mass outflow rate \Mdotout\ of a constant velocity spherical or (multi-)conical outflow  can be calculated from the \Ha\ luminosity of the outflow ($L_{\rm H\alpha, out}$) as follows:
\begin{multline}
\dot{M}_{\rm out} \, {\rm (M_\odot \, yr^{-1}}) = 33 \left( \frac{1000~{\rm cm}^{-3}}{n_e} \right) \left( \frac{v_{\rm out}}{1000~{\rm km~s}^{-1}} \right) \\ \times \left( \frac{1~{\rm kpc}}{R_{\rm out}} \right) \left( \frac{L_{\rm H\alpha, out}}{10^{43}~{\rm erg~s^{-1}}} \right) \label{eqn:mdot_out}
\end{multline}
\rout\ is the radial extent of the outflow, \vout\ is the outflow velocity and $n_e$ is the local electron density of the ionized gas in the outflow \citep{Genzel11, Newman12_406690}\footnote{Equation \ref{eqn:mdot_out} assumes that the outflowing gas is photoionized. If the ionized gas in the outflow is primarily collisionally excited and has a temperature of \mbox{$\sim$~2 $\times$ 10$^4$ K}, the mass outflow rates would scale by a factor of $\sim$~0.6 \citep[see Appendix B of][]{Genzel11}.}. The mass loading factor $\eta$ is defined as $\eta$~=~\Mdotout/SFR$_{\rm best}$.
\end{sloppypar}

The biggest uncertainty in the calculation of \Mdotout\ and $\eta$ is the electron density, which is extremely challenging to constrain for AGN-driven outflows at z$\sim$2. In principle, the electron density of the ionized gas in the outflow can be measured from the \SII$\lambda$6716/\SII$\lambda$6731 ratio in the outflow component \citep{Osterbrock06}. However, the \SII\ lines are weak compared to \Ha\ and \NII, and become strongly blended in the presence of a broad outflow component, making it very difficult to constrain a two component decomposition of the \SII\ line profile. \citet{NMFS19} fit two kinematic components to a high S/N stacked spectrum of 30 AGN-driven outflows at \mbox{0.6~$< z < $~2.6}, and found that the average electron density of the outflowing material is $\sim$~1000~cm$^{-3}$. Their stack includes two of the three galaxies in our sample. Several recent studies of AGN-driven outflows at low and intermediate redshift have found similarly high densities in the outflowing gas \citep[e.g.][]{Perna17, Kakkad18, Husemann19, Shimizu19}. We therefore adopt $n_e$~=~1000~cm$^{-3}$.

\subsection{Extinction Correction}\label{subsec:extcor}
The \Ha\ emission from the outflowing gas is attenuated by dust along the line of sight to the nuclear regions of the galaxies. To calculate the intrinsic mass outflow rates, we need to correct the observed \Ha\ fluxes for extinction. 

Our observations cover both the \Ha\ and \Hb\ emission lines. When \Hb\ is detected, the best method for correcting emission line luminosities for extinction is to use the Balmer decrement, which directly probes the attenuation along the line of sight to the nebular line emitting regions. The \Hb\ line is too weak to robustly separate into disk and outflow components, and therefore we adopt the integrated Balmer decrement. The theoretical unattenuated Balmer decrement for Case B recombination at \mbox{T~=~10$^4$~K} is 2.86 \citep{Osterbrock06}. However, this value can increase to 3.1 in the presence of an AGN \citep{Gaskell84}. Therefore, we adopt an intrinsic Balmer decrement of 3.1. We assume the nebular extinction follows the \citet{Cardelli89} curve. 

\begin{sloppypar}
In cases where the Balmer decrement cannot be measured, we correct for extinction using the global continuum $A_V$ of the galaxy, and account for extra attenuation of the nebular emission using the empirical formula presented by \citet{Wuyts13} (\mbox{A$_{\rm H\alpha}$ = 1.9 A$_{\rm stars}$ - 0.15 A$_{\rm stars}^2$}). The empirical correction assumes that the extinction follows the \citet{Calzetti00} curve, but the functional form was chosen to produce the best agreement between the \Ha\ and UV~+~IR SFRs for SFGs at 0.7~$<$~z~$<$~1.5 (i.e. \mbox{A$_{\rm H\alpha}$~$\sim$~-2.5 log$_{\rm 10}$ (SFR$_{\rm H\alpha}$/SFR$_{\rm UV+IR}$)}), and therefore the derived A$_{\rm H\alpha}$ should be largely independent of the chosen attenuation curve. 
\end{sloppypar}

We note that in both cases we assume that the outflow emission and galaxy nebular emission experience similar attenuation, which may not be the case if the two components have different spatial distributions. Higher S/N and/or space-based observations of the \Hb\ line are required to calculate the Balmer decrements of the galaxy and outflow components separately.

\begin{figure*}
\centering
\hspace{2cm} \includegraphics[scale=0.7, clip = True, trim = 10 30 5 0]{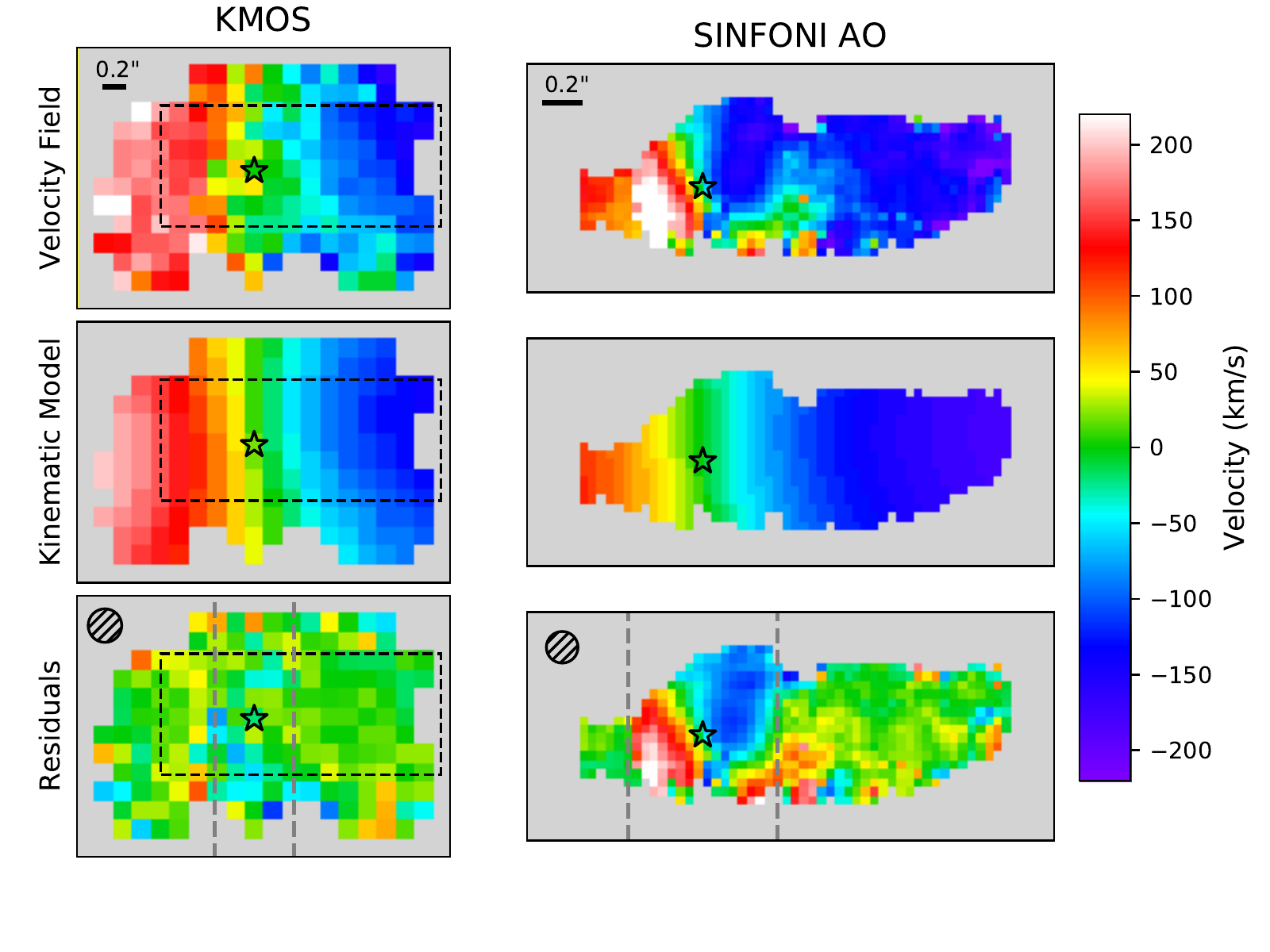} \\ \includegraphics[scale = 0.76, clip=True,trim = 10 140 40 0]{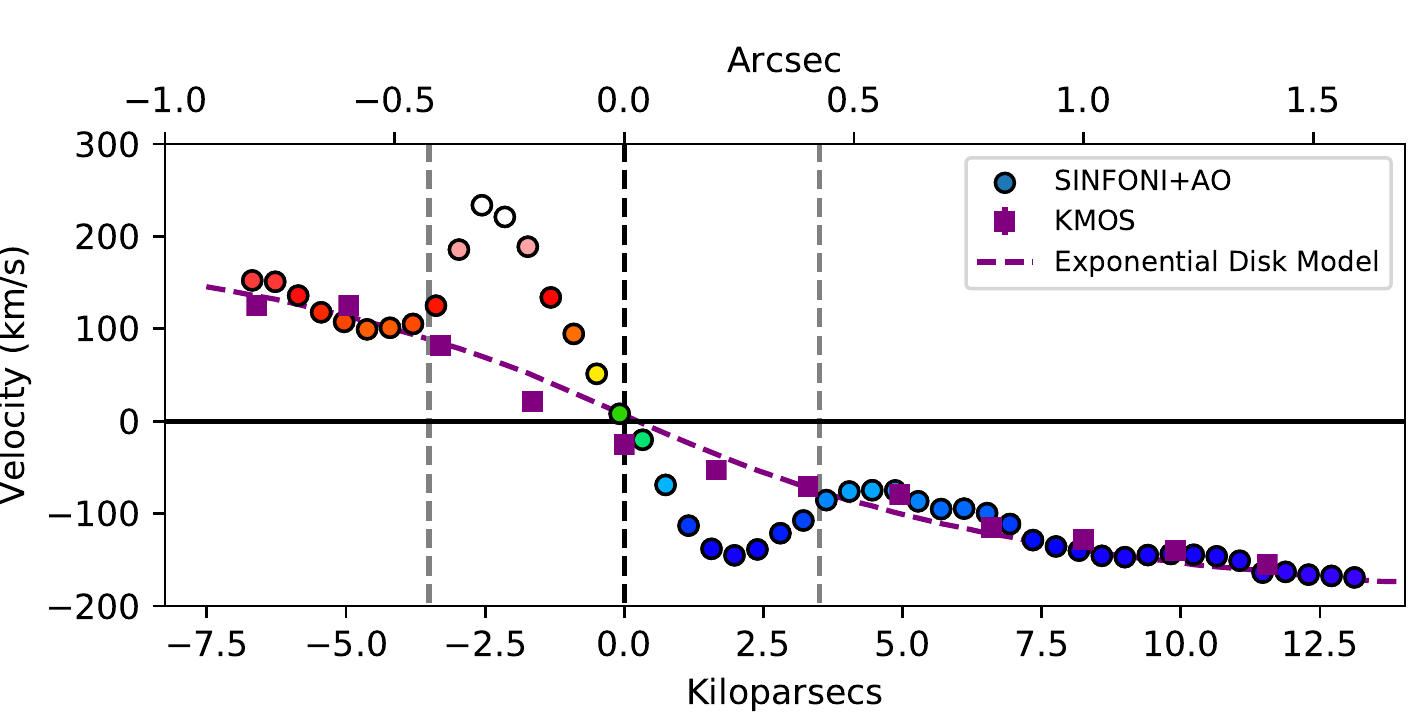}
\caption{\underline{Top}: Summary of the 2D kinematic modelling for K20-ID5. The first row of panels shows the velocity fields measured from the KMOS (left) and SINFONI-AO (right) data. The SINFONI-AO data covers a sub-region of the KMOS data, indicated by the black dashed rectangle. The black stars indicate the kinematic center of the galaxy. The second row shows the exponential disk model fit to the KMOS velocity field, mock observed at the spatial resolution and sampling of both datasets. The third row shows the residuals after subtracting the best fit model from the measured velocity fields. A strong nuclear velocity gradient is visible in the SINFONI-AO residuals, across the region bounded by the grey dashed lines. \underline{Bottom}: 1D velocity profiles extracted along the kinematic major axis \mbox{(PA = -84.5$^\circ$)}. The colored dots trace the velocity profile measured from single component fits to SINFONI-AO spectra, where the colors represent the velocity on the same color scale used for the velocity field maps. The purple squares indicate the velocity profile measured from the KMOS data, and the purple dashed line shows the profile of the exponential disk model fit to the KMOS velocity field. \label{fig:id5_kinematic_model}}
\end{figure*}

\section{K20-ID5: A Powerful Galaxy Scale Outflow}\label{sec:id5}

\subsection{Velocity Field and Kinematic Modelling}\label{subsec:id5_modelling}
Our SINFONI-AO data reveal previously unresolved kinematic structures in K20-ID5, providing key insights into the nature of the line emission in the nuclear region of the galaxy. The kinematic properties of \mbox{K20-ID5} are summarized in Figure \ref{fig:id5_kinematic_model}.

\begin{figure*}
\centering
\includegraphics[scale=0.6, clip = True, trim = 80 30 00 20]{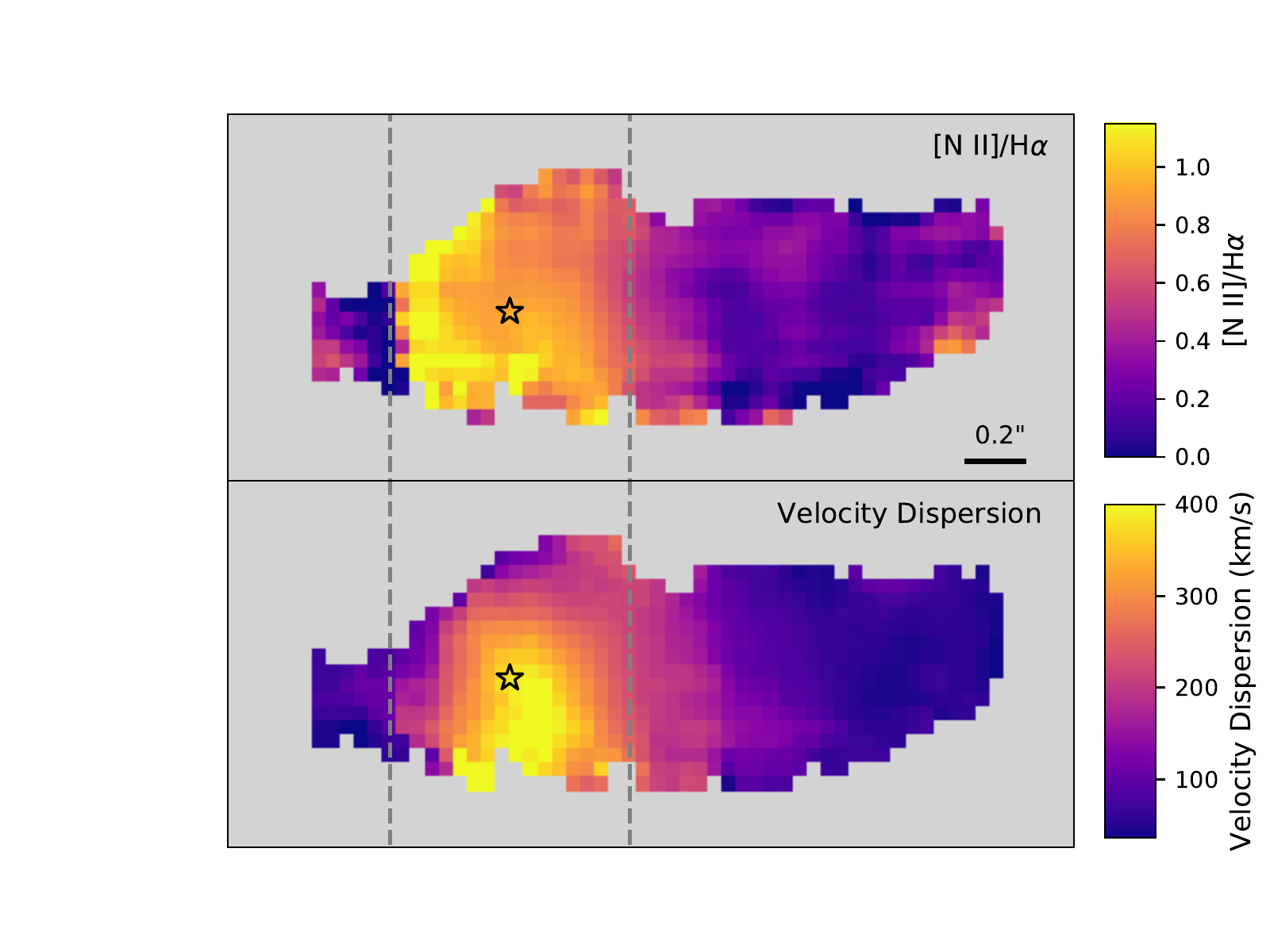} \includegraphics[scale=0.6, clip = True, trim = 0 30 0 0]{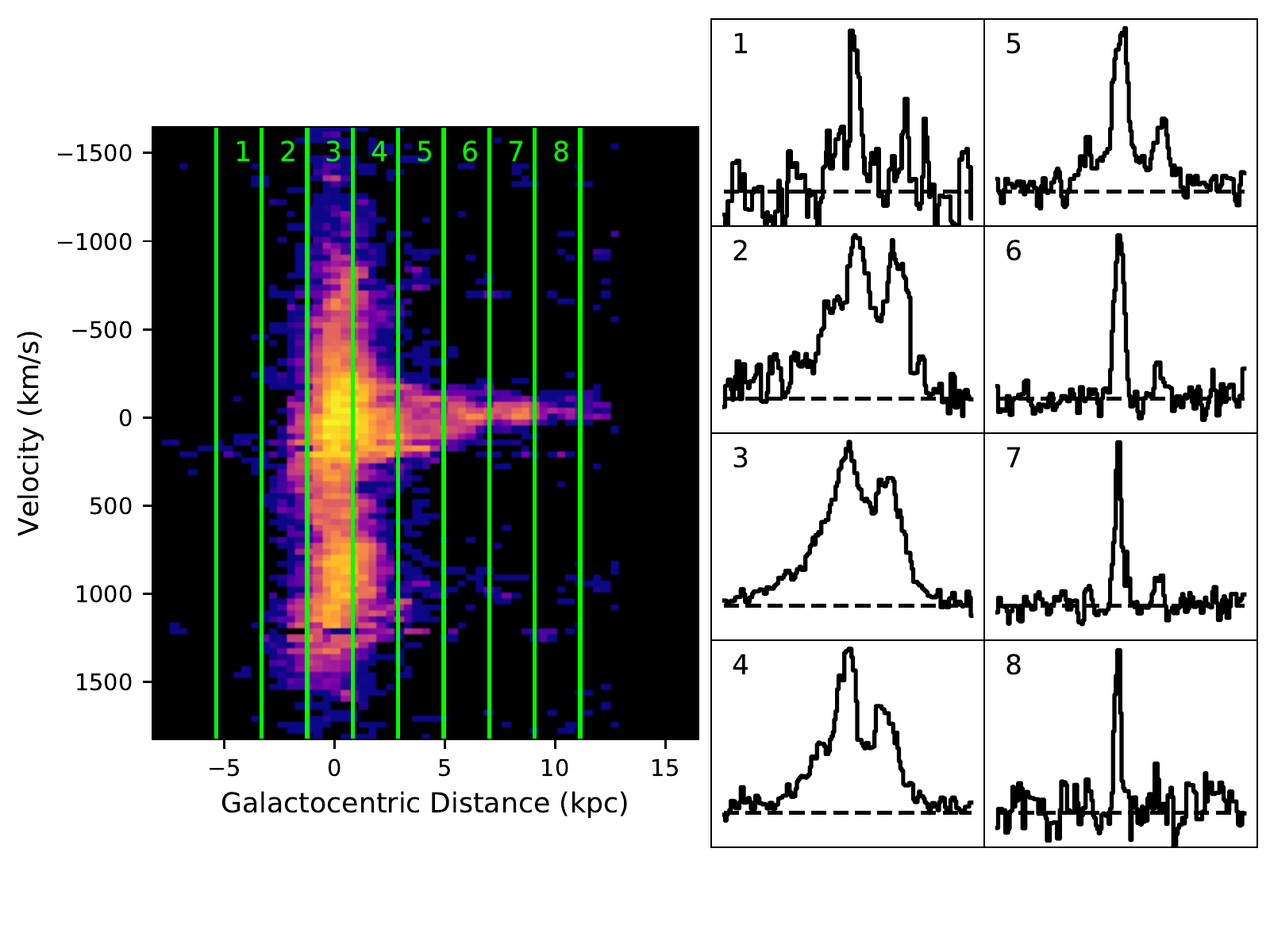}\\
\caption{Left: Maps of the \NIIHa\ ratio (top) and single component velocity dispersion (bottom) for K20-ID5. The grey dashed lines delineate the region where strong velocity residuals are observed (see Figure \ref{fig:id5_kinematic_model}). Right: Position-velocity (p-v) diagram for K20-ID5 (constructed by collapsing the cube along the N-S direction), and 1D spectra extracted in eight slices along the p-v diagram. \label{fig:id5_pv_diagram}}
\end{figure*}

The first row of panels shows the velocity fields measured from the KMOS and SINFONI-AO data. At the lower resolution and coarser spatial sampling of the KMOS data, the velocity field looks quite regular, but the higher resolution \mbox{SINFONI-AO} data reveal a twist and a steepening of the velocity gradient in the central 0.4''. The bottom panel of Figure \ref{fig:id5_kinematic_model} shows the 1D velocity profile, extracted in 5$\times$5 spaxel apertures along the kinematic major axis of the galaxy \mbox{(PA = -84.5$^\circ$)}. The colored dots trace the velocity of the ionized gas as a function of galactocentric distance in the SINFONI-AO data, and the purple squares show the velocity profile extracted from the KMOS data. The velocities are derived from single component Gaussian fits to the \Ha\ and \NII\ lines, but we verified that very similar profiles are recovered when performing two component fitting to account for the presence of the broad emission component in the nuclear region. Therefore, the SINFONI-AO data reveal irregular narrow component kinematics in the nuclear region of \mbox{K20-ID5}.

We exploit the regularity of the KMOS velocity field to model the kinematics of the extended disk. We model the velocity field using \textsc{Dysmalpy}, a python implementation of the dynamical fitting code \textsc{Dysmal} \citep{Cresci09, Davies11, Wuyts16, Uebler18}. \textsc{Dysmal} is a forward modeling code which builds a model for the mass distribution (including one or more of a disk, bulge and dark matter halo), produces a mock datacube with a given spatial and spectral sampling, convolves the datacube with the given line spread function and spatial PSF, produces model velocity and velocity dispersion fields, and compares the model to the data. We fit the KMOS velocity field using an exponential disk model with the mass, effective radius, inclination and position angle of the disk as free parameters. We do not fit the velocity dispersion field because it is heavily influenced by the outflow in the central regions.

The best fit model for the extended velocity field of K20-ID5 is shown in the second row of Figure \ref{fig:id5_kinematic_model}, at the resolution/sampling of the KMOS data in the left column and at the resolution/sampling of the SINFONI-AO data in the right column. The third row of panels shows the residuals after subtracting the best fit model from the measured velocity field for each of the datasets. The best fit model reproduces the KMOS velocity field very well, leaving only small amplitude residuals (less than 55 \kms\ in 90\% of spaxels) with no clear spatial structure. On the other hand, the discontinuity in the SINFONI-AO velocity field is visible as a strong velocity gradient in the residuals, going from -120~\kms\ to +240~\kms\ at an angle of $\sim$~25$^\circ$ to the major axis of the disk.

The twist in the narrow component kinematics could trace either a misaligned core or an outflow. In order to determine which of these is more likely, we examine how the line profiles and the single component \NIIHa\ ratio and velocity dispersion differ between the nuclear region and the extended disk of the galaxy. The left hand panels of Figure \ref{fig:id5_pv_diagram} show maps of the single component \NIIHa\ ratio (top) and velocity dispersion (bottom) from the SINFONI-AO data. The \NIIHa\ ratio is elevated in the nucleus compared to the extended disk region, and more interestingly shows a sharp boundary between the nucleus and the extended disk on the eastern (receding) side of the galaxy. This sharp boundary coincides with an abrupt change in the magnitude of the velocity residuals. The velocity dispersion is also elevated in the nuclear region, but decreases gradually with increasing galactocentric distance.

The right hand panels of Figure \ref{fig:id5_pv_diagram} show a position-velocity (p-v) diagram extracted along the E-W direction (i.e. the p-v diagram and the maps on the left side of the figure share the same $x$-axis), and spectra extracted in eight slices which are numbered on both the p-v diagram and the corresponding panels. Slices 1, 6, 7 and 8 trace the regularly rotating extended disk of the galaxy and show narrow line emission and low \NIIHa\ ratios. Slices 2-4 trace the kinematically anomalous nuclear region, and show strong broad emission and high \NIIHa\ ratios. Slice 5 traces the western edge of the nuclear region where the broad outflow emission is still visible but is overpowered by narrow emission from the galaxy disk. The \NIIHa\ ratio is intermediate between the nuclear and extended disk regions, but the narrow lines follow the velocity field of the extended disk. 

The kinematically anomalous nuclear region of K20-ID5 is characterized by broad line profiles and high \NIIHa\ ratios, whereas the regularly rotating regions at larger radii are dominated by narrow disk emission with much lower \NIIHa\ ratios. This provides strong circumstantial evidence that the residual nuclear velocity gradient is tracing the outflow.

The nuclear velocity gradient reflects the kinematics of the narrow line peaks, leading to the conclusion that both the narrow and broad line emission in the nuclear region of K20-ID5 must be primarily associated with outflowing material. In other words, the outflow has a non-Gaussian line-of-sight velocity distribution which can be approximated by the superposition of a narrow and a broad Gaussian component. 3D biconical outflow models suggest that outflow emission line profiles can have non-Gaussian shapes, and may resemble the superposition of a narrow and a broad component depending on the outflow velocity and geometry \citep[e.g.][]{Bae16}. AGN driven outflows with non-Gaussian line profiles and/or requiring multiple Gaussian components have been observed both in the local universe \citep[e.g.][]{Fischer18, Shimizu18, DaviesRic20} and at high redshift \citep[e.g.][]{Liu13b, Vayner17}. There are also many examples of AGN host galaxies with outflow-dominated narrow line region kinematics \citep[e.g.][]{Fischer13, Liu13b}, similar to what we observe in K20-ID5.

\begin{figure*}
\centering
\includegraphics[scale=1, clip = True, trim = 10 110 210 0]{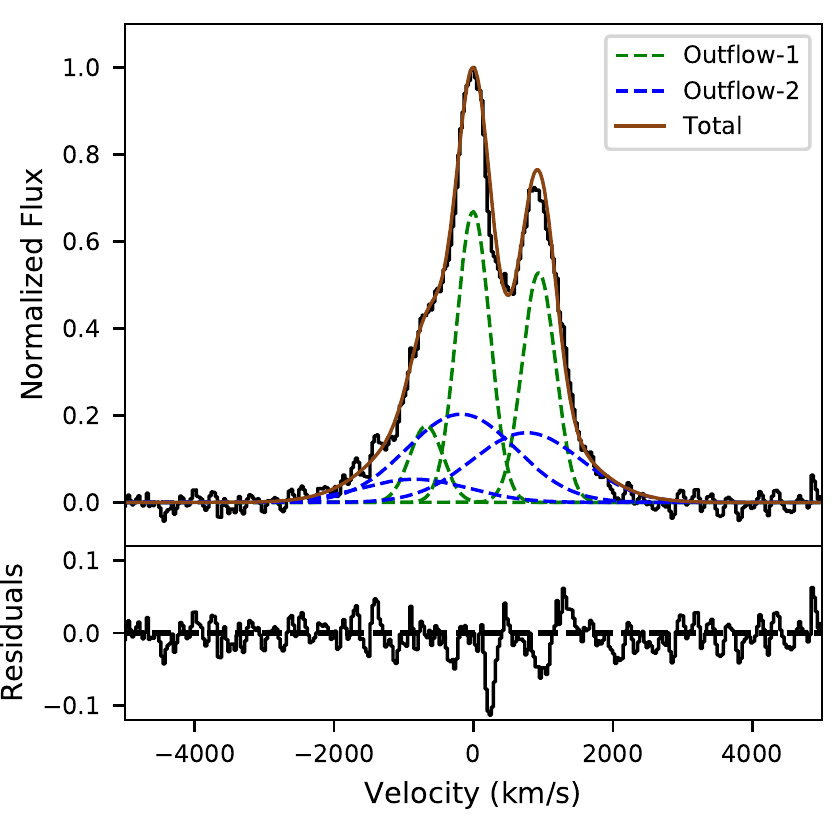} \includegraphics[scale=1, clip = True, trim = 0 110 210 0]{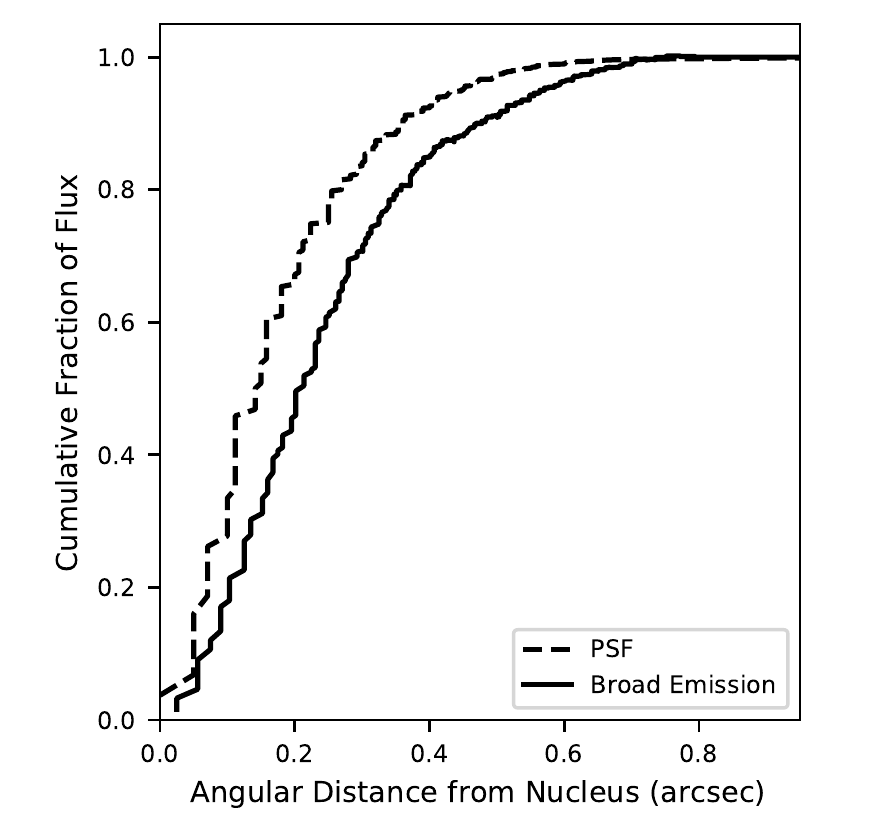}
\caption{(Left) Deep SINFONI-AO + KMOS nuclear spectrum of K20-ID5 (black) with the best fit two component Gaussian model over-plotted. The green and blue curves represent the narrow and broad components respectively, and the brown curve represents the total best fit line profile. In this spectrum, both components are attributed to the outflow. The bottom panel shows the fit residuals. (Right) Curve of growth of the SINFONI-AO PSF (dotted) and the broad emission (solid) as a function of radius. \label{fig:id5_nuc_spec_fit}}
\end{figure*}

We note that the clear distinction between the kinematics and line ratios of the nuclear outflow and the extended gas in K20-ID5 confirm that the extended gas is tracing the rotating disk of the galaxy rather than a galaxy-scale outflow (see discussion in \citealt{Loiacono19}).

\subsection{Outflow Velocity}\label{subsec:id5_outflow_kinem}
We measure the properties of the nuclear outflow from a deep K band spectrum extracted over the kinematically anomalous nuclear region of K20-ID5, where the outflow dominates the line emission (see Section \ref{subsec:id5_modelling}). We combine the SINFONI-AO and KMOS datasets for a total on-source integration time of 31 hours. The spectrum is shown in the left hand panel of Figure \ref{fig:id5_nuc_spec_fit}.

Based on the FWZP of the \NII+\Ha\ complex, we measure an outflow velocity of 1410~$\pm$~56~\kms.\footnote{We note that our outflow velocity is a factor of $\sim$3 higher than the outflow velocity reported in \citet{Genzel14}. This difference likely arises because 1) the dataset used in this work has a factor of seven longer integration time, significantly improving the S/N in the high velocity wings of the line profile, 2) the \citet{Genzel14} measurement was based on the KMOS nuclear spectrum, and the broad component is significantly more prominent in the SINFONI-AO nuclear spectrum because the higher spatial resolution allows the outflow-dominated nuclear region to be better separated from the surrounding disk-dominated regions, and 3) they adopted a more conservative definition of the outflow velocity.} Our best-fit exponential disk model yields a circular velocity of $\sim$~240~\kms, corresponding to a halo escape velocity of $\sim$~720~\kms\ (assuming \mbox{$v_{\rm escape} \sim 3~v_{\rm circ}$}; \citealt{Weiner09}). The outflow velocity is a factor of two larger than the escape velocity, suggesting that a significant fraction of the outflowing material could be ejected from the halo.

\subsection{Outflow Energetics}\label{subsec:id5_energetics}
The mass outflow rate is derived from the \Ha\ luminosity of the outflow, as described in Section \ref{subsec:mdot}. Our analysis in Section \ref{subsec:id5_modelling} revealed that the vast majority of the (narrow and broad) line emission in the nuclear region of K20-ID5 is associated with the outflow. Therefore, we calculate the outflow properties assuming that 100\% of the nuclear \Ha\ flux originates from the outflow (``outflow'' model). 

Even though the outflow dominates the line emission, it is necessary to fit two Gaussian components to recover the shape of the line profiles, as shown in the left hand panel of Figure \ref{fig:id5_nuc_spec_fit}. We sum the \Ha\ fluxes of the two components to obtain the total \Ha\ flux. We do not use the seeing limited SINFONI H band data to constrain the emission line fitting, because the nuclear \OIII\ emission is strongly attenuated (see Figure \ref{fig:id5_footprints}) and we do not find any evidence for a broad or blueshifted component in the \OIII\ line profile.

\begin{table*}[]
\begin{nscenter}
\caption{Derived outflow parameters.}
\begin{tabular}{lcccc}
\hline Galaxy & \multicolumn{2}{c}{K20-ID5}   &  COS4-11337  & J0901 \\ \hline
Model Type & \textbf{Outflow} & Galaxy~+~Outflow & \textbf{Galaxy~+~Outflow} & \textbf{Galaxy~+~Outflow} \\ 
(1) & (2) & (3) & (4) & (5) \\ \hline \hline
a) \rout\ (kpc) & \multicolumn{2}{c}{1.0~$\pm$~0.2} & 0.9~$\pm$~0.2 & 0.47~$\pm$~0.07 \\
b) \vout\ (\kms) & \multicolumn{2}{c}{1410~$\pm$~56} & 1459~$\pm$~66 & 650~$\pm$~46 \\
c) \Mdotout\ (M$_\odot$ yr$^{-1}$) & 262~$\pm$~76 & 103~$\pm$~30 & 61~$\pm$~6 & 25~$\pm$~8 \\
d) $\eta$ (= \Mdotout/SFR$_{\rm best}$) & 0.78~$\pm$~0.23 & 0.31~$\pm$~0.09 & 0.15~$\pm$~0.03 & 0.12~$\pm$~0.04 \\ \hline
\end{tabular}
\end{nscenter}
\tablecomments{For COS4-11337 and J0901 we consider only a \mbox{Galaxy + Outflow} model, whilst for \mbox{K20-ID5} we also consider a model where all of the nuclear line emission is associated with the outflow (Outflow). Boldface font indicates the fiducial model for each galaxy. The rows are as follows: \mbox{a) Half} light radial extent of the outflow emission. b) Outflow velocity, defined as \mbox{(FWZP$_{\rm [N II]+H\alpha}$ - 1600)/2}. \mbox{c) Mass} outflow rate, calculated using Equation \ref{eqn:mdot_out}. d) Mass loading factor.}\label{table:outflow_properties}
\end{table*}

There is a significant amount of dust attenuating the nebular line emission from the nuclear region of \mbox{K20-ID5} (see Section \ref{subsec:id5_properties}). We use the SINFONI H band data to measure the \Hb\ flux, and find \mbox{\Ha/\Hb~=~8.0~$\pm$~0.7}, corresponding to an A$_V$ of 2.7. This is significantly larger than the global continuum $A_V$ of 1.3, but is consistent with the results of \citet{Loiacono19} who measured a global Balmer decrement of 8.3~$\pm$~1.8, and \citet{Scholtz20} who measured a nuclear Balmer decrement of 8.7$^{+2.3}_{-1.8}$. We use the measured Balmer decrement to correct the \Ha\ luminosity of the outflow for extinction.

The outflow is well resolved (see right hand panel of Figure \ref{fig:id5_nuc_spec_fit}), and has a PSF-deconvolved HWHM of 1.0~$\pm$~0.2~kpc. Combining all these quantities, we measure a mass outflow rate of \mbox{262~$\pm$~76} \mbox{M$_\odot$~yr$^{-1}$}, corresponding to a mass loading factor of \mbox{$\eta$ = 0.78~$\pm$~0.23}.

For comparison, we also calculate the outflowing mass using just the \Ha\ luminosity of the broad component (``galaxy~+~outflow'' model). In this case, we measure \mbox{\Mdotout~=~103~$\pm$~30} M$_\odot$~yr$^{-1}$ and \mbox{$\eta$ = 0.31~$\pm$~0.09}. The outflow parameters for the outflow only (fiducial) and galaxy~+~outflow models are listed in Columns 2 and 3 of Table \ref{table:outflow_properties}, respectively.

\subsection{Outflow Geometry and Velocity Structure}\label{subsec:id5_geometry}
The detection of a velocity gradient across the nucleus of K20-ID5 makes it possible to place some constraints on the geometry of the outflowing material. The velocity difference between the approaching and receding sides of the outflow is \mbox{$\Delta v \sim$~360~\kms} (see Figure \ref{fig:id5_kinematic_model}), which is significantly smaller than the outflow velocity (1410~\kms). This suggests that the outflow is quasi-spherical, because a large opening angle ($\ga$~60$^\circ$) is required to produce a large range of projected outflow velocities at every radius, while maintaining a similar average velocity across the entire outflow \citep[see e.g. models in][]{Liu13b}. The small $\Delta v$ could also be produced by a collimated outflow almost perpendicular to the line of sight, but in a collimated outflow there would only be a small range of velocities at each radius, and therefore this scenario cannot account for the large observed line width.

It is also possible to place some constraints on the velocity profile of the outflowing material. There is no evidence for any radial variation in the FWZP of the \NII+\Ha\ complex, suggesting that the outflow velocity is approximately constant out to the maximum radius at which it is detected (at least $\sim$~5~kpc; see Figure \ref{fig:id5_pv_diagram}).

\section{COS4-11337: An Outflow in a Galaxy Pair}\label{sec:cos4_11337}

\subsection{Outflow Velocity and Energetics}\label{subsec:cos4_11337_outflow_kinem}
We measure the properties of the outflow from \mbox{COS4-11337} by fitting the KMOS H and K band nuclear spectra as a superposition of two Gaussian components. We use the KMOS K band data in favour of the SINFONI-AO K band data because the KMOS K band observations have 3.3 times the integration time and therefore a factor of $\sim$1.8 higher S/N, and allow us to perform a robust galaxy + outflow decomposition of the nuclear line profile which is not possible using the SINFONI data alone. Although \mbox{COS4-11363} and COS4-11337 are partially blended in the KMOS data, the SINFONI-AO data indicate that the line emission from COS4-11363 is weak and is confined to the nucleus of the galaxy (see Figure \ref{fig:cos4_11363_footprints}). Therefore, the contribution of \mbox{COS4-11363} to the nuclear spectrum of COS4-11337 should be negligible.

\begin{figure*}
\centering
\includegraphics[scale=1,clip=True,trim = 0 180 0 0]{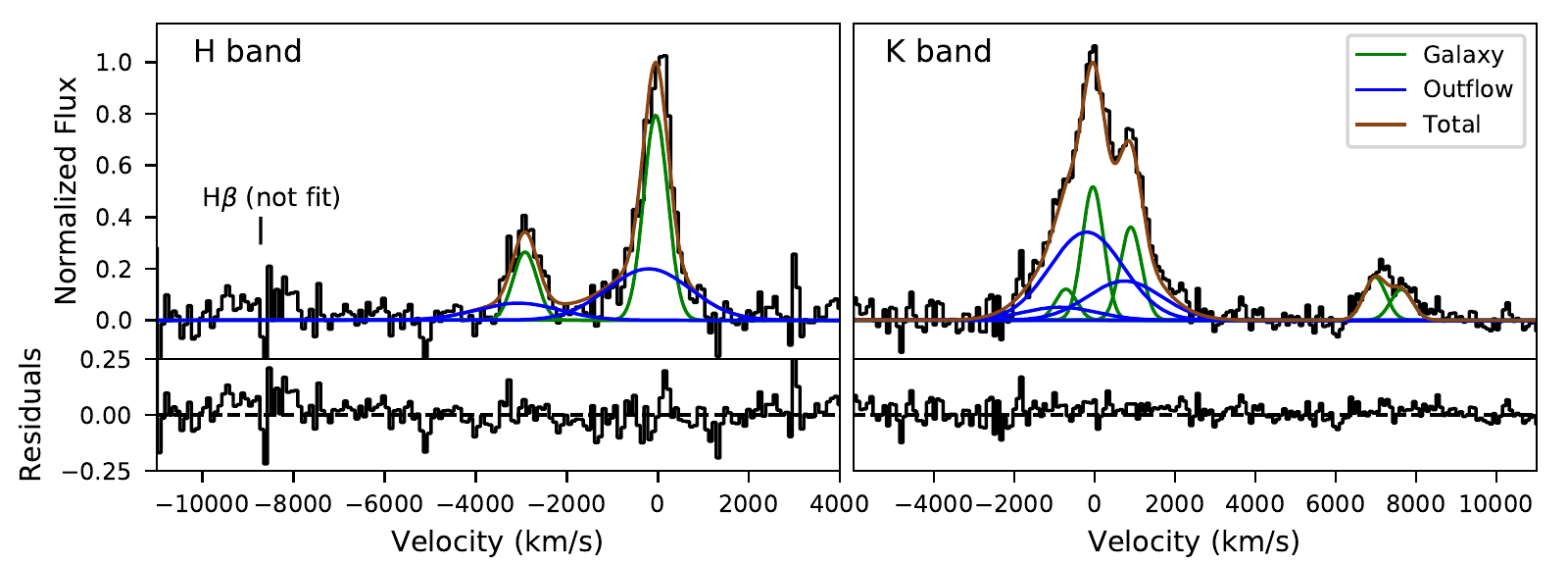}
\caption{KMOS H and K band nuclear spectra of COS4-11337, plotted in independent normalized flux units. The green and blue curves indicate the best fit narrow (galaxy) and broad (outflow) components for a two component Gaussian fit, respectively. The brown curve represents the sum of the narrow and broad components, and the lower panels show the fit residuals in each band. \label{fig:cos4_11337_nuc_spec_fit}}
\end{figure*}

\begin{figure}
\centering
\includegraphics[scale=1,clip=True,trim = 10 145 235 40]{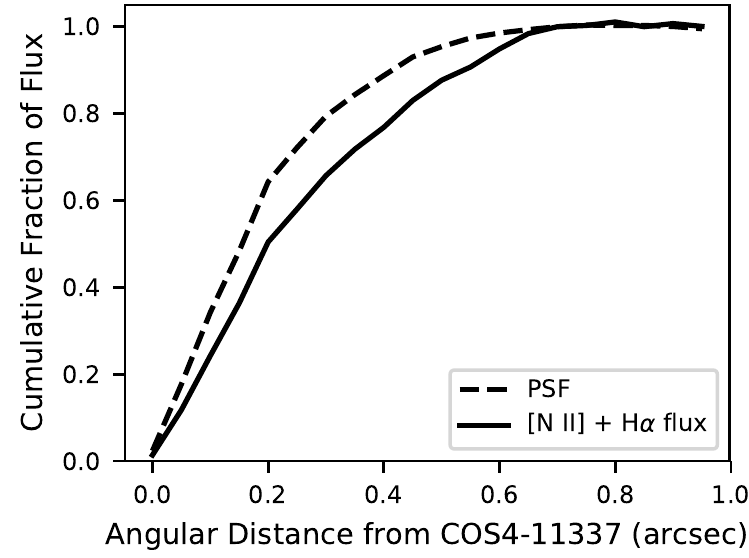}
\caption{Curve of growth of the SINFONI-AO PSF (dashed) and \NII+\Ha\ flux (solid) as a function of distance from the center of COS4-11337. \label{fig:cos4_11337_psf}}
\end{figure}

The two component fit to the spectrum of \mbox{COS4-11337} is shown in Figure \ref{fig:cos4_11337_nuc_spec_fit}. The K band spectrum shows very broad wings, revealing the presence of a fast outflow. Based on the FWZP of the \NII+\Ha\ complex, we measure an outflow velocity of 1459~$\pm$~66~\kms. \mbox{COS4-11337} has a circular velocity of 150~$\pm$~60~\kms\ \citep{Wisnioski18}, corresponding to an escape velocity of $\sim$~450~\kms. The outflow velocity is more than a factor of three larger than the escape velocity, indicating again that a significant fraction of the outflowing material could potentially be expelled from the host halo.

\subsection{Outflow Energetics}
The redshift of this system places the \Hb\ line at a wavelength with bad skyline residuals (see Figure \ref{fig:cos4_11337_nuc_spec_fit}), and as a result we cannot derive a reliable Balmer decrement for COS4-11337. Therefore, we correct the \Ha\ luminosity of the outflow for extinction using the global A$_V$, as described in Section \ref{subsec:extcor}. The outflow is resolved (see Figure \ref{fig:cos4_11337_psf}) and has a PSF-deconvolved HWHM of 0.9~$\pm$~0.2 kpc. We find a mass outflow rate of 61~$\pm$~6~M$_\odot$~yr$^{-1}$ and a mass loading factor of $\eta$~=~0.15~$\pm$~0.03. The derived outflow parameters are listed in Column 4 of Table \ref{table:outflow_properties}. 

\begin{figure*}
\centering
\includegraphics[scale=0.8,clip=True,trim = 10 150 10 20]{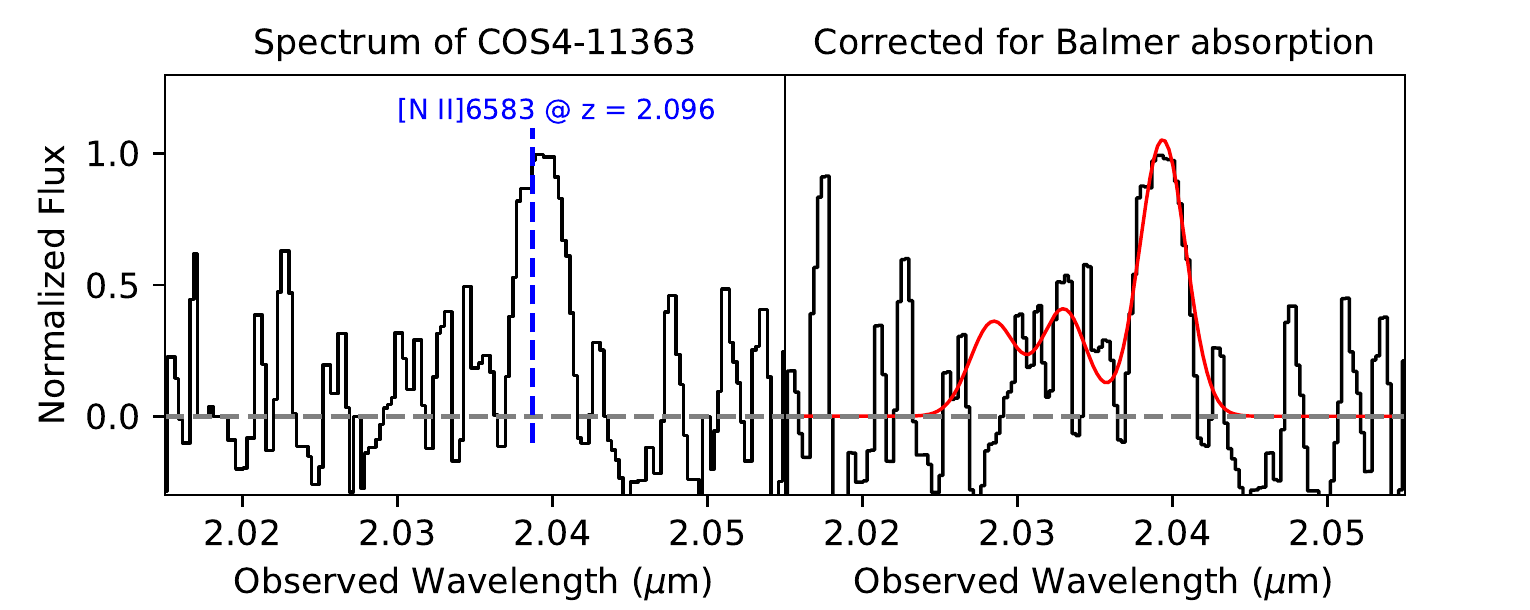}
\caption{(Left) SINFONI-AO K band nuclear spectrum for COS4-11363. One emission line is detected at $\lambda\sim$~2.039~$\mu$m. This emission line lies close to the wavelength of the \NII\ line at the redshift of COS4-11337, which is indicated by the blue dashed line. (Right) Spectrum corrected for stellar absorption, with the fit to the \NII+\Ha\ line emission shown in red. \label{fig:cos4_11363_sinfao_spec}}
\end{figure*}

\begin{figure*}
\centering
\includegraphics[scale=0.84,clip=True,trim = 10 170 240 20]{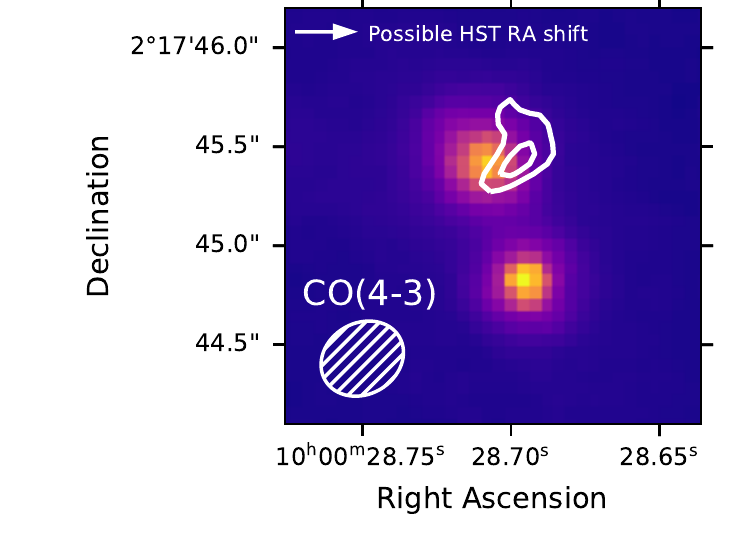} \includegraphics[scale=0.74,clip=True,trim = 0 150 20 10]{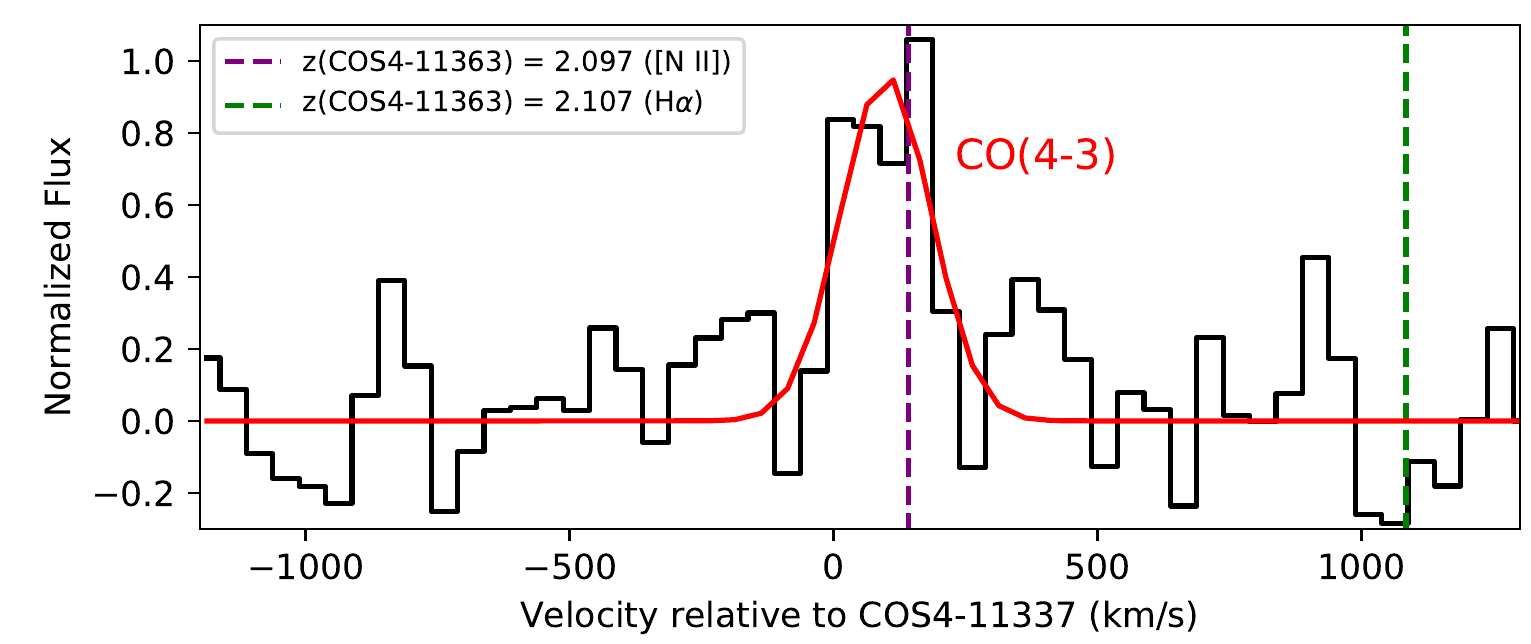}
\caption{Left: HST F160W image of the COS4-11337/11363 system, with 2$\sigma$ and 3$\sigma$ contours of the CO(4-3) emission overlaid. The CO(4-3) emission is clearly associated with COS4-11363. The offset between the CO and F160W centroids may be an artefact of a small offset in the HST RAs. Right: CO(4-3) spectrum extracted over the region where the line is detected, plotted as a function of velocity relative to COS4-11337. The red line indicates the best fit to the line emission and is centered at $z$~=~2.097. The purple and green lines indicate the possible redshifts of COS4-11363 from the SINFONI-AO data (see Figure \ref{fig:cos4_11363_sinfao_spec}), assuming the detected emission line is \NII$\lambda$6583 at $z$~=~2.097, or \Ha\ at $z$~=~2.107, respectively. The CO(4-3) detection is strongly in favour of the detected line being \NII$\lambda$6583. \label{fig:cos4_11363_co43}}
\end{figure*}

\subsection{Redshift of COS4-11363}
COS4-11337 and COS4-11363 are resolved and clearly separated in the SINFONI-AO K band data, allowing us to extract and analyse the spectrum of \mbox{COS4-11363}, which is shown in the left hand panel of Figure \ref{fig:cos4_11363_sinfao_spec}. Only one emission line is detected. It is relatively narrow and lies close to the observed wavelength of \NII$\lambda$6583 in COS4-11337, which is indicated by the blue dashed line. The single emission line in the spectrum of COS4-11363 could trace either \NII$\lambda$6583 at $z$~=~2.097, in which case the d$v$ between the galaxies would be $\sim$~140~\kms, or \Ha\ at $z$~=~2.107, in which case the d$v$ would be $\sim$~1100~\kms. The 3D-HST grism redshift is $z_{\rm grism}$~=~2.103, in between the two possible spectroscopic redshifts.

We break the redshift degeneracy by utilising archival ALMA observations. COS4-11337/11363 was observed for 90 minutes in Band 4 as part of program 2016.1.00726.S (PI: A. Man). The observations cover the CO(4-3) line, and have a spatial resolution of 0.39'' which is sufficient to resolve the two galaxies. The left hand panel of Figure \ref{fig:cos4_11363_co43} shows the F160W image of the system, with contours of the \mbox{CO(4-3)} emission (at levels of 2$\sigma$ and 3$\sigma$) overlaid. Despite the relatively short integration time, CO(4-3) emission is detected near the nucleus of COS4-11363\footnote{We note that the offset between the CO and F160W centroids may be the result of a small offset in the HST RAs. We compared the HST and Gaia positions of the two Gaia stars in 3D-HST COSMOS tile 12, and found that the HST RAs were lower by 0.14'' and 0.28''. There were no significant DEC offsets. The white arrow in Figure \ref{fig:cos4_11363_co43} indicates how the HST data would shift relative to the ALMA data if the RAs were to increase by 0.28''.}. The right hand panel shows the CO(4-3) spectrum associated with the peak of the emission. The spectrum is plotted as a function of velocity offset from COS4-11337. The purple and green dashed lines indicate where CO(4-3) would fall if the redshift of COS4-11363 were $z$~=~2.097 or $z$~=~2.107, respectively. The ALMA data clearly favour the $z$~=~2.097 scenario, indicating that the line detected in the SINFONI-AO K band spectrum is \NII$\lambda$6583, and that the $\Delta v$ between COS4-11363 and COS4-11337 is $\leq$~150~\kms.

\subsection{Nature of the Line Emission in COS4-11363}\label{subsec:line_emission_11363}
The detection of \NII$\lambda$6583 emission without strong \Ha\ emission indicates that the \NIIHa\ ratio in this galaxy must be significantly higher than observed in normal star forming galaxies. The average \NIIHa\ ratio for a pure star forming, \mbox{log(M$_*$/M$_\odot$)~=~11.1} galaxy on the mass-metallicity relation at $z$~=~2.1 is \mbox{\NIIHa~$\sim$~0.4} (based on the mass-redshift-metallicity parametrisation in \citet{Tacconi18} and the \NIIHa-metallicity calibration from \citealt{Pettini04}). However, to determine the intrinsic \NIIHa\ ratio in the nucleus of COS4-11363, we must account for the fact that the \Ha\ emission line coincides with a deep photospheric absorption feature in the spectrum of A stars. The best fit SED models for COS4-11363 support the presence of strong Balmer absorption features, regardless of whether we adopt a truncated or exponentially declining star formation history. We scale the best fit SED to match the continuum level of COS4-11363, and subtract this scaled best-fit SED from the observed spectrum (shown in the left hand panel of Figure \ref{fig:cos4_11363_sinfao_spec}) to obtain a pure emission line spectrum (right panel of Figure \ref{fig:cos4_11363_sinfao_spec}). We fit the \NII\ and \Ha\ lines in this emission line spectrum as single Gaussians, using the same fitting algorithm with the same parameter restrictions as described for our multi-component fitting process (see Section \ref{subsec:separating_outflow_emission}). We measure an \NII/\Ha\ ratio of 2.6~$\pm$~0.4 - a factor of 6.5 higher than expected for a pure star forming galaxy. 

The high \NIIHa\ ratio indicates that the line-emitting gas must be collisionally excited and/or ionized by sources other than young stars. We measure an \Ha\ equivalent width of 5.8~$\pm$~2.5 \AA, which exceeds the maximum value of 3 to be consistent with ionization by evolved stellar populations \citep[e.g.][]{CidFernandes11, Belfiore16}. We therefore suggest that the line emission is most likely to be powered by either shock excitation or AGN activity.

\begin{figure}
\centering
\includegraphics[scale=1.0,clip=True,trim = 0 160 30 10]{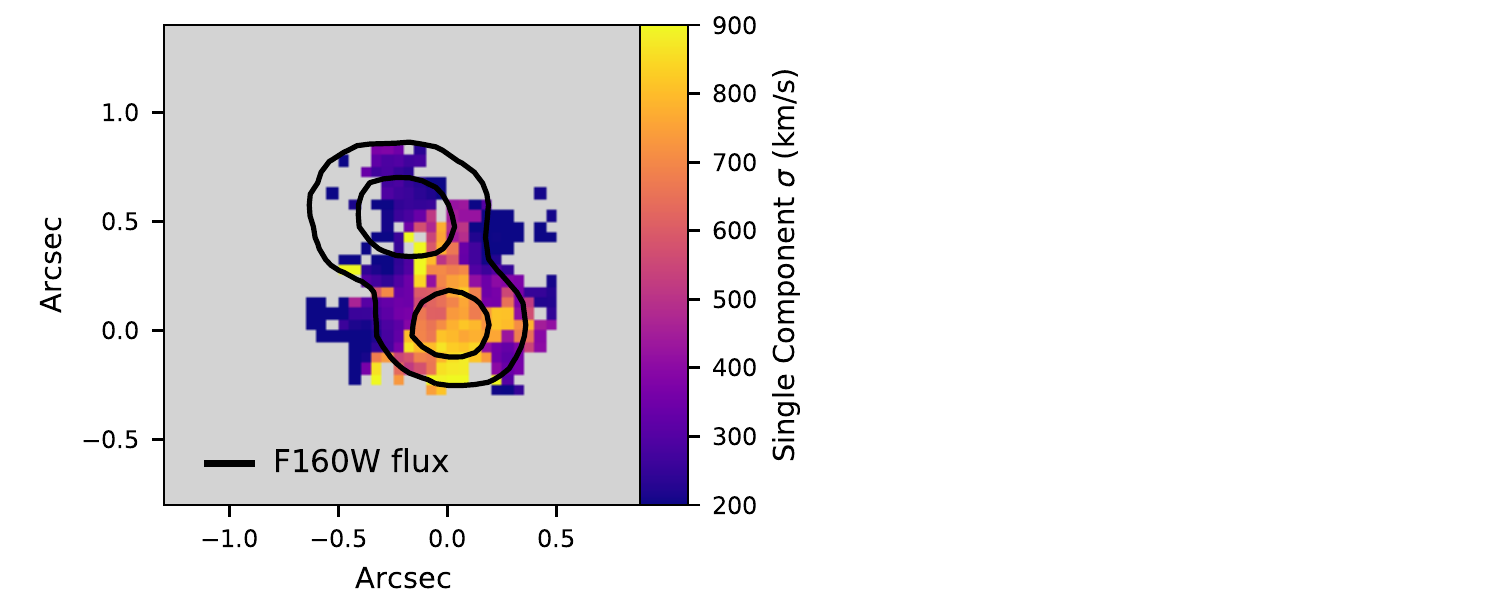}
\caption{Map of the single component velocity dispersion across the \mbox{COS4-11337/11363} system. The velocity dispersion peaks at the nucleus of COS4-11337 but remains elevated in the region between the two galaxies, suggesting that the outflow from COS4-11337 may be propagating towards COS4-11363. \label{fig:cos4_11363_vdisp_map}}
\end{figure}

The strongest evidence for the source of the line emission comes from the velocity dispersion map, shown in Figure \ref{fig:cos4_11363_vdisp_map}. The velocity dispersion peaks at $\sim$~800~\kms\ at the nucleus of COS4-11337, where the outflow is launched. However, it remains elevated above 500~\kms\ along the entire region connecting COS4-11363 and COS4-11337, before dropping to $\sim$~250~\kms\ at the nucleus of COS4-11363. This suggests that the outflow from \mbox{COS4-11337} may be propagating towards its companion. Based on the outflow velocity and the projected separation between the galaxies, the travel time between the nuclei is $\sim$~4~Myr. If the outflow collided with the ISM of COS4-11363, it is likely to have driven large scale shocks, producing the high observed \NIIHa\ ratios. This is a potential example of AGN feedback acting on both galactic scales (by transporting mass and energy in the outflow) and circumgalactic scales (by driving shocks through the ISM of the companion galaxy). 

Tidal torques are likely to be an additional source of shock excitation in both COS4-11363 and COS4-11337. In the local universe, interacting/merging systems show prominent line emission from shock excited gas with a typical FWHM of 250-500~\kms\ \citep[e.g.][]{Monreal-Ibero06, Farage10, Rich11, Rich15}. This FWHM is similar to the width of the line emission from COS4-11363, but a factor of $\sim$4 narrower than the broad line emission from COS4-11337, suggesting that an outflow is most likely required to explain the kinematics of the gas in COS4-11337. However, we cannot rule out a scenario where the shock excitation in \mbox{COS4-11363} is purely induced by the interaction.

\begin{figure}
\centering
\includegraphics[scale=1.0,clip=True,trim = 10 110 150 0]{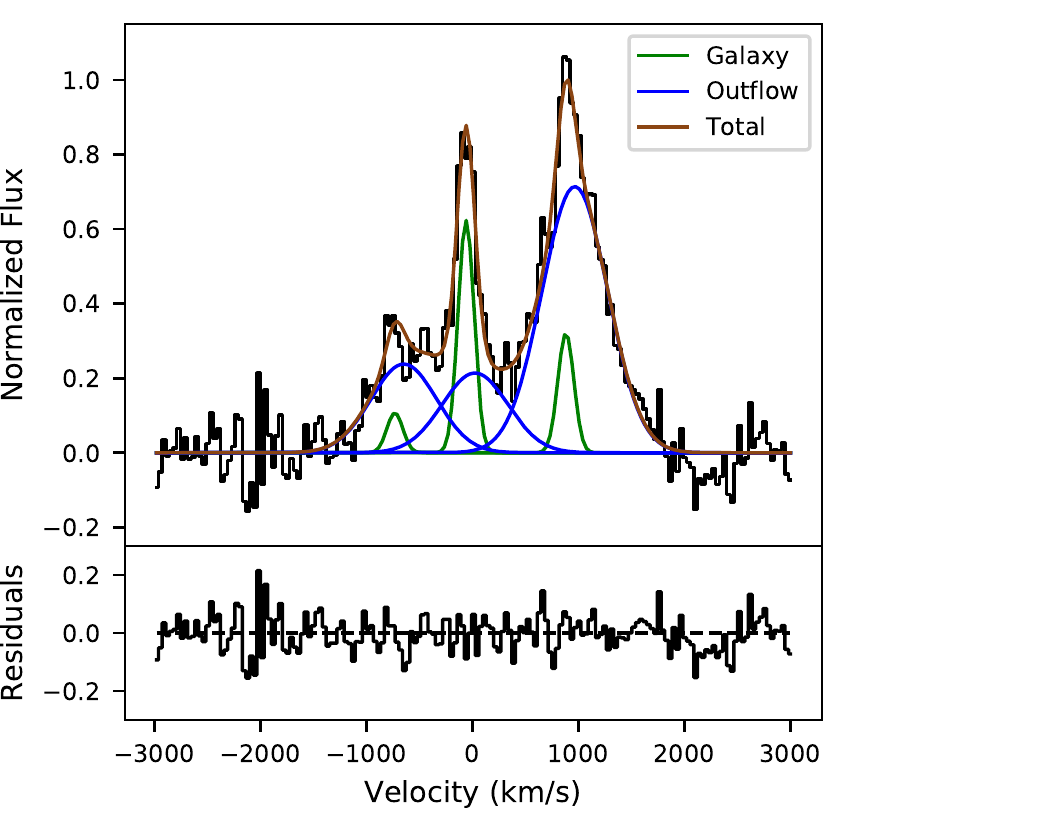}
\caption{Source plane nuclear spectrum of J0901 (black), with the best two component fit over-plotted. The narrow (galaxy) and broad (outflow) components are shown in green and blue, respectively, and the total fit is shown in brown. The bottom panel shows the fit residuals. \label{fig:j0901_nuc_fit}}
\end{figure}

It is interesting to speculate on the possible impact that the outflow from COS4-11337 may have had on the star formation activity of COS4-11363. The 3D-HST grism spectrum and the best fit SED for \mbox{COS4-11363} both support the presence of a prominent Balmer break, and the SED fitting favours a rapid decrease in the SFR over the last hundred Myr. From the \Ha\ flux, we measure an instantaneous SFR of $<$~2~M$_\odot$~yr$^{-1}$, which places the galaxy two orders of magnitude below the main sequence SFR. The detection of CO associated with this galaxy indicates that the recent truncation of the SF activity was not simply the consequence of an exhausted gas reservoir. The quenching could plausibly have been triggered by the outflow from \mbox{COS4-11337} ploughing into the ISM of \mbox{COS4-11363}, driving large scale shocks and preventing the gas from collapsing to form stars. However, the galaxy-galaxy interaction is also likely to have had a significant impact on the SF activity in this system. Deep spectroscopy covering the region around 4000\AA\ will assist to more accurately constrain the star formation history and evolution of this galaxy.

\begin{figure*}
\centering
\includegraphics[scale=0.9, clip = True, trim = 0 160 10 10]{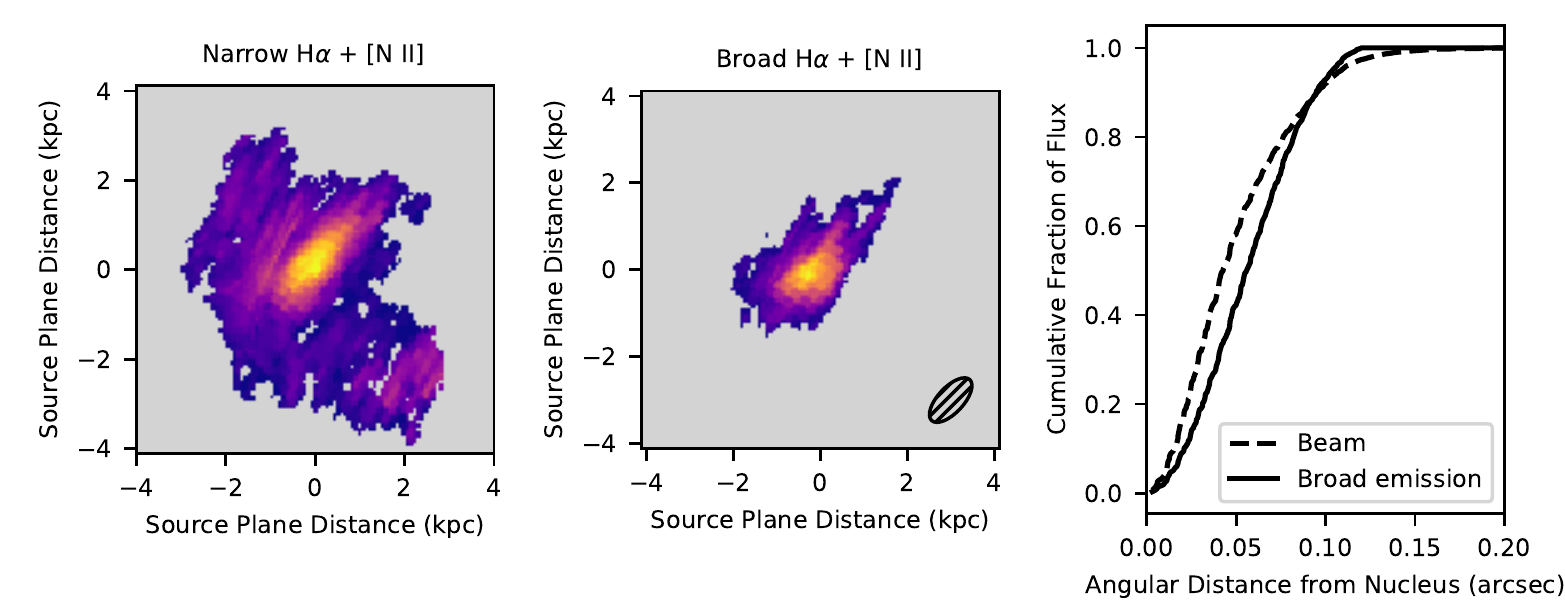} 
\caption{Left and middle: Source plane maps of the narrow and broad \Ha\ and \NII\ emission across J0901, at a pixel scale of 0.005''. The hatched ellipse at the bottom right of the middle panel indicates the approximate shape and size of the source plane PSF. Right: Curves of growth for the beam and the broad component. The broad component is slightly more extended than the beam, indicating that it is marginally resolved. \label{fig:j0901_broad_narrow_maps}}
\end{figure*}

\section{J0901: A Centrally Confined Outflow}\label{sec:j0901}
\subsection{Outflow Velocity}\label{subsec:j0901_spec_decomp}
The nuclear spectrum of J0901 was extracted from the source plane datacube (which was created by applying the source plane reconstruction to each spectral channel individually), and is shown in Figure \ref{fig:j0901_nuc_fit}. The \NII\ line is significantly broader than the \Ha\ line, and our two component fitting reveals that this is because the \Ha\ emission peak is dominated by the galaxy component whereas the \NII\ emission peak is dominated by the broader outflow component. From the FWZP of the \NII+\Ha\ complex, we measure an outflow velocity of 650~$\pm$~46~\kms. This is a factor of $\sim$2 smaller than the outflow velocities measured for \mbox{K20-ID5} and \mbox{COS4-11337}. Kinematic modelling indicates that J0901 has an inclination-corrected circular velocity of $\sim$~260~\kms\ \citep{Rhoads14,Sharon19}, corresponding to an escape velocity of 780~\kms. The outflow velocity is approximately 85\% of the escape velocity, and therefore the majority of the outflowing material detected in our data is unlikely to be able to escape from the galaxy halo. 

\subsection{Outflow Extent}
Figure \ref{fig:j0901_broad_narrow_maps} shows source plane maps of the narrow galaxy emission (left) and the broad outflow emission (middle) over the region covered by the SINFONI-AO data. The approximate shape of the source plane PSF at the location of the nucleus is indicated by the hatched ellipse at the bottom right of the middle panel. Both the galaxy and outflow emission show a single peak in the nucleus, unlike the image plane \NII\ map (white contours in the right hand panel of Figure \ref{fig:j0901_footprints}) in which two clear peaks are visible. Our lens modelling reveals that the secondary peak in the image plane flux distribution is a re-image of the northern side of the nucleus. This result is confirmed by \citet{Sharon19}, who independently developed a lens model for the J0901 lensing cluster in parallel with our team.

The narrow emission is extended and traces the disk of the galaxy. We note that this figure only shows the central region of the galaxy covered by our AO data, and the true extent of the disk is significantly larger. In contrast, the broad outflow emission is very centrally concentrated. The right hand panel of Figure \ref{fig:j0901_broad_narrow_maps} compares the curves of growth for the outflow emission and the PSF. The source plane spatial resolution of the SINFONI-AO data at the nucleus is \mbox{122~$\times$~56 mas} (FWHM), corresponding to a circularized FWHM of 83~mas and a physical resolution of 680~pc.  Despite the factor of two improvement in spatial resolution due to the lensing, the outflow emission is only slightly more extended than the PSF, indicating that it is marginally resolved. The de-convolved Gaussian HWHM of the outflow emission is 470~$\pm$~70~pc - a factor of $\sim$~2 smaller than the outflows from K20-ID5 and COS4-11337.

\subsection{Outflow Energetics}\label{subsec:j0901_energetics}
The outflowing mass is calculated from the source plane \Ha\ luminosity of the outflow component. We robustly detect the \Hb\ line in our LUCI long slit spectrum of J0901, and therefore we correct the \Ha\ luminosity for extinction using the Balmer decrement. We measure a Balmer decrement of \mbox{\Ha/\Hb\ = 6.8~$\pm$~0.6}, corresponding to $A_V$~=~2.2. We find a mass outflow rate of \mbox{25~$\pm$~8 M$_\odot$ yr$^{-1}$}, and a mass loading factor of \mbox{$\eta$ = 0.12~$\pm$~0.04} (listed in Column 5 of Table \ref{table:outflow_properties}). 

\section{Discussion}\label{sec:discussion}

\subsection{Outflow Driving Mechanisms}\label{subsec:energetics}
Although all three of our galaxies host AGN, the observed ionized gas outflows could plausibly be driven by SF, AGN activity, or both. Outflows can be energy conserving or momentum conserving depending on how fast the hot wind material cools as it interacts with the ISM \citep[e.g.][]{King10, King15}.

\begin{table*}[]
\begin{nscenter}
\caption{Outflow kinetic powers and momentum rates, and comparison to the bolometric luminosities of the AGN and the young stars.}
\begin{tabular}{lcccc}
\hline Galaxy & \multicolumn{2}{c}{K20-ID5}   &  COS4-11337  & J0901 \\ \hline
Model Type & \textbf{Outflow} & Galaxy~+~Outflow & \textbf{Galaxy~+~Outflow} & \textbf{Galaxy~+~Outflow} \\ 
(1) & (2) & (3) & (4) & (5) \\ \hline \hline
log($\dot{E}_{\rm out, n_e = 1000}$/(erg s$^{-1}$)) & 44.2~$\pm$~0.1 & 43.8~$\pm$~0.1 & 43.6~$\pm$~0.1 & 42.5~$\pm$~0.2 \\
log($\dot{p}_{\rm out, n_e = 1000}$/dyn) & 36.4~$\pm$~0.1 & 36.0~$\pm$~0.1 & 35.7~$\pm$~0.1 & 35.0~$\pm$~0.1 \\
$\dot{E}_{\rm out, n_e = 1000}$/$L_{\rm AGN}$ & 
(4.0~$\pm$~1.2)$\times$10$^{-2}$ & (1.6~$\pm$~0.5)$\times$10$^{-2}$ & (2.8~$\pm$~0.3)$\times$10$^{-3}$ & (1.5~$\pm$~0.9)$\times$10$^{-4}$ \\
$\dot{p}_{\rm out, n_e = 1000}$/($L_{\rm AGN}$/c) & 17~$\pm$~5 & 6.8~$\pm$~2.0 & 1.1~$\pm$~0.1 & 0.14~$\pm$~0.08 \\ 
$\dot{E}_{\rm out, n_e = 380}$/$L_{\rm SF,best}$ & (3.4~$\pm$~1.0)$\times$10$^{-2}$ & (1.3~$\pm$~0.4)$\times$10$^{-2}$ & (7.1~$\pm$~1.2)$\times$10$^{-3}$ & (1.1~$\pm$~0.5)$\times$10$^{-3}$ \\
$\dot{p}_{\rm out, n_e = 380}$/($L_{\rm SF,best}$/c) & 14~$\pm$~4 & 5.6~$\pm$~1.7 & 2.9~$\pm$~0.5 & 1.0~$\pm$~0.4 \\ \hline
\end{tabular}
\end{nscenter}
\tablecomments{The columns are the same as in Table \ref{table:outflow_properties}. When comparing to L$_{\rm AGN}$ we adopt n$_e$~=~1000~cm$^{-3}$, and when comparing to L$_{\rm SF}$ we adopt n$_e$~=~380~cm$^{-3}$, reflecting the different ionized gas densities of AGN-driven and SF-driven outflows at $z\sim$~2 (see discussion in Section \ref{subsec:energetics}).}\label{table:outflow_energy_momentum}
\end{table*}

The AGN ionizing radiation field injects momentum into the surrounding material, driving a hot wind with a momentum rate of \mbox{$\dot{p}_{\rm wind}$ $\sim$~L$_{\rm AGN}$/c}. If the kinetic energy of the wind couples efficiently to the ISM, it can drive a galaxy scale energy conserving outflow with kinetic power \mbox{\Edotout\ $\lesssim$~0.05~L$_{\rm AGN}$} and momentum rate \mbox{\pdotout\ $\lesssim$ (5-20)~L$_{\rm AGN}$/c} \citep{Faucher12, Zubovas12}. On the other hand, if the wind kinetic energy is efficiently radiated away, the result is a smaller scale momentum conserving outflow with \mbox{\Edotout\ $\lesssim$~10$^{-3}$~L$_{\rm AGN}$} and \mbox{\pdotout\ $\lesssim$~L$_{\rm AGN}$/c} \citep{King10, King15}. 

Supernovae and stellar winds deposit energy at a rate of \mbox{\Edotout\ $\lesssim$~10$^{-3}$~L$_{\rm SF}$} \citep{Murray05}. Momentum is deposited through both radiation pressure from massive stars ($\dot{p}_{\rm rad}$~$\sim$~L$_{\rm SF}$/c) and supernova explosions. The initial momentum of the ejecta from a single supernova is \mbox{$p_{\rm i, SN}$ $\sim$~3000~M$_\odot$ \kms}, and assuming one supernova per 100 years per solar mass of SF, this corresponds to a total momentum injection rate by supernovae of \mbox{$\dot{p}_{\rm i, SN}$ $\sim$~0.7~L$_{\rm SF}$/c} \citep{Murray05}. However, if the cooling time of the supernova ejecta is sufficiently long, the hot wind can sweep up a significant amount of ISM, and the final momentum rate of the outflow can be a factor of $\sim$~10 larger than the initial wind momentum rate ($\dot{p}_{\rm f, SN}$ $\sim$~6~L$_{\rm SF}$/c; \citealt{Kim15}). SF driven outflows have been observed to propagate to distances of a few kiloparsecs at $z\sim$~2 \citep[e.g.][]{Newman12_406690, Newman12_global, Davies19}. 

We calculate the kinetic powers and momentum rates of the outflows in our three galaxies and compare them to L$_{\rm SF}$ and L$_{\rm AGN}$ to determine if the outflows are driven by SF or AGN activity and if they are momentum or energy conserving (see Table \ref{table:outflow_energy_momentum}).

In this comparison it is important to account for the 1/n$_e$ dependence of the mass outflow rate. In our calculations we have adopted n$_e$~=~1000~cm$^{-3}$, under the assumption that the outflows are AGN-driven. However, the typical electron density of the ionized gas in SF-driven outflows at $z\sim$~2 is 380~cm$^{-3}$ \citep{NMFS19}, a factor of 2.6 lower. Therefore, when calculating \Edotout/L$_{\rm SF}$ and \pdotout/(L$_{\rm SF}$/c), we multiply the mass outflow rates by a factor of 2.6. It is also important to consider that L$_{\rm SF}$ represents the \textit{global} bolometric output of the young stars, but the AGN-driven outflows are launched from the nuclear regions of the galaxies, and therefore only some fraction of L$_{\rm SF}$ will be available to drive the outflows. 

The kinetic powers of the outflows from K20-ID5 and COS4-11337 are too large for the outflows to be SF-driven. The kinetic power of the K20-ID5 outflow is 0.04($\pm$0.01)~$\times$~L$_{\rm AGN}$ and the momentum rate is 17($\pm$5)~$\times$~L$_{\rm AGN}/c$, suggesting that the outflow is energy conserving. However, we note that the energy and momentum ratios vary inversely with the AGN luminosity which is poorly constrained due to the probable high column densities towards the nucleus (see Section \ref{subsec:id5_properties}). The coupling between the AGN radiation field and the outflow is less efficient in COS4-11337, which has an outflow kinetic power of \mbox{2.8($\pm$0.3)~$\times$~10$^{-3}$~L$_{\rm AGN}$} and a momentum rate of 1.1($\pm$0.1)~$\times$~L$_{\rm AGN}/c$. The inefficient coupling suggests that the outflow could plausibly be momentum driven. However, the outflow extends to a distance of $\sim$~5 kpc (see Section \ref{subsec:line_emission_11363} and Figure \ref{fig:cos4_11363_vdisp_map}), which is more in favour of the energy driving scenario.

The J0901 outflow has a smaller kinetic power than the outflows from K20-ID5 and COS4-11337, primarily due to the lower outflow velocity. The outflow has a kinetic power of \mbox{1.1($\pm$0.5)~$\times$~10$^{-3}$~L$_{\rm SF}$} and a momentum rate of 1.0($\pm$0.4)~$\times$~L$_{\rm SF}$/$c$, and could therefore potentially be SF-driven. However, the spatial distribution of the \Ha\ emission suggests that only $\gtrsim$~1/3 of the SF is occuring within the outflow region, in which case the bolometric luminosity of the young stars would likely be insufficient to power the observed outflow. In addition, the outflow emission has a very high \NIIHa\ ratio (3.3; see Figure \ref{fig:j0901_nuc_fit}), which is typical of AGN-driven outflows but is not observed in SF-driven outflows at $z\sim$~2 \citep[e.g.][]{Newman12_406690, Davies19, Genzel14, NMFS19}. Therefore, we suggest that the outflow is most likely to be AGN-driven, although we cannot rule out a significant contribution from the SF. Only a very small fraction of the energy released by the AGN needs to couple to the ISM in order to drive the observed outflow. Given the very low coupling efficiency and fact that the outflow is confined to the circumnuclear regions, it seems likely that the outflow in J0901 is a momentum conserving AGN-driven outflow.

\subsection{Extents of AGN-Driven Outflows}
An important ingredient in our understanding of how AGN-driven outflows interact with their host galaxies is accurate measurements of the radial extents of AGN-driven outflows. In all three of our systems we find that the majority of the outflow emission is concentrated within approximately the central kiloparsec, consistent with the findings of \citet{NMFS14}. However, with the exception of J0901, the outflows are not confined to the nuclear regions, but extend well beyond the effective radii of the galaxies. In K20-ID5 the outflow propagates at an approximately constant velocity to a radius of at least 5~kpc (see Section \ref{subsec:id5_geometry}). In \mbox{COS4-11337}, we see compelling evidence that the outflow has travelled beyond the galaxy, and is shock heating the ISM in its companion galaxy COS4-11363 (see Section \ref{subsec:line_emission_11363}). Other studies of luminous AGN at similar redshifts have found evidence that ionized outflows can propagate 5-10 kpc from the galaxy nuclei \citep[e.g.][]{Nesvadba06, Nesvadba08, Harrison12, Cresci15,Brusa18, HerreraCamus19b}. Put together, these results suggest that AGN-driven outflows have steep luminosity profiles, with a luminous core component in the central kpc and a faint tail extending to several kpc, which may reflect a decrease in the surface brightness and/or density of the outflowing material as it expands out from the galaxy nuclei \citep[e.g.][]{Kakkad18}.

However, the detection of a confined \mbox{(r$_e$~=~470~$\pm$~70~pc)} outflow in J0901 emphasises that not all AGN-driven outflows extend beyond the nuclear region. \citet{Fischer19} reported another hundred parsec scale, $\sim$~500~\kms\ AGN-driven outflow in the lensed galaxy SGAS J003341.5+024217 (SGAS 0033+02 hereafter) at $z$~=~2.39. The outflows in both J0901 and SGAS 0033+02 are so compact that they would not be resolved without gravitational lensing. \citet{Fischer19} state that the outflow in SGAS 0033+02 would probably not be detectable if the system was not lensed, because the outflow emission would be overpowered by emission from the galaxy. The J0901 outflow is clearly visible even in an 0.6'' (source plane) diameter aperture, primarily because the large \NIIHa\ ratio in the broad component increases the contrast between the broad component and the surrounding continuum. However, it is possible that confined nuclear outflows are present but undetected in a non-negligible fraction of massive galaxies at $z\sim$~2.

The similarity in the AGN luminosities measured for COS4-11337 and J0901 suggests that the outflow extent is not determined by the \textit{current} AGN luminosity, although we re-iterate that the uncertainty on the AGN luminosity of J0901 is relatively large. The AGN in J0901 and \mbox{SGAS 0033+02} could potentially have `switched on' relatively recently, so that the outflows have not yet had time to propagate beyond the circumnuclear regions. Alternatively, the AGN accretion energy may not couple efficiently to the gas in the nuclear regions (as appears to be the case for J0901), giving the outflows insufficient energy to propagate to larger radii. In their current states, these outflows are unable to directly impact gas on kiloparsec scales. However, the outflows are depositing a significant amount of kinetic energy within a few hundred parsecs of the galaxy nuclei. This deposition of energy will increase the amount of turbulence in the circumnuclear regions, and if the turbulent pressure becomes large enough, the star-forming gas could become stabilized against collapse \citep[e.g.][]{Guillard12, Alatalo15a}. Therefore, these small, lower velocity outflows could potentially lead to a reduction of the star formation efficiency in the nuclear regions of their host galaxies. Further studies of confined outflows will assist to better characterize this unique class of objects.

\subsection{Mass and Energy Budget of AGN-Driven Outflows}\label{subsec:multiphase_outflows}
The near infrared spectroscopic data analyzed in this paper probe only the ionized gas phase of the AGN-driven outflows in K20-ID5, COS4-11337 and J0901. However, galaxy scale outflows are intrinsically multi-phase, and contain not only warm ionized gas, but also cooler molecular and  atomic gas, and hotter X-ray emitting gas. Multi-phase observations of outflows in local AGN host galaxies suggest that on galaxy scales, the molecular and neutral phases dominate the outflow mass and the mass outflow rate, but the ionized gas has a higher outflow velocity \citep[e.g.][]{Rupke13, Veilleux13, Fiore17, Fluetsch19, Husemann19, HerreraCamus19b, Shimizu19}. In two quasar-driven outflows at $z\sim$~1.5, the molecular phase has a factor of $\sim$2-5 higher mass outflow rate than the ionized phase but a factor of $\sim$2-4 lower outflow velocity \citep[e.g.][]{Vayner17, Brusa18}. For a typical star forming galaxy at $z\sim$~2, \citet{HerreraCamus19} found that the molecular outflow rate is a factor of $\sim$5 higher than the ionized outflow rate.

Simulations predict that the hot ($\sim$~10$^7$~K) phase should carry at least as much mass as the cooler gas phases \citep[e.g.][]{Nelson19}, but so far the majority of the observational constraints come from studies of \mbox{X-ray} Broad Absorption Line (BAL) winds and Ultra-Fast Outflows (UFOs) on very small spatial scales, and these appear to have mass outflow rates similar to or lower than those of ionized gas outflows \citep[e.g.][]{Feruglio15, Fiore17, Tombesi17}.  

\begin{figure*}
\centering
\includegraphics[scale=0.7,clip=True,trim = 10 10 10 10]{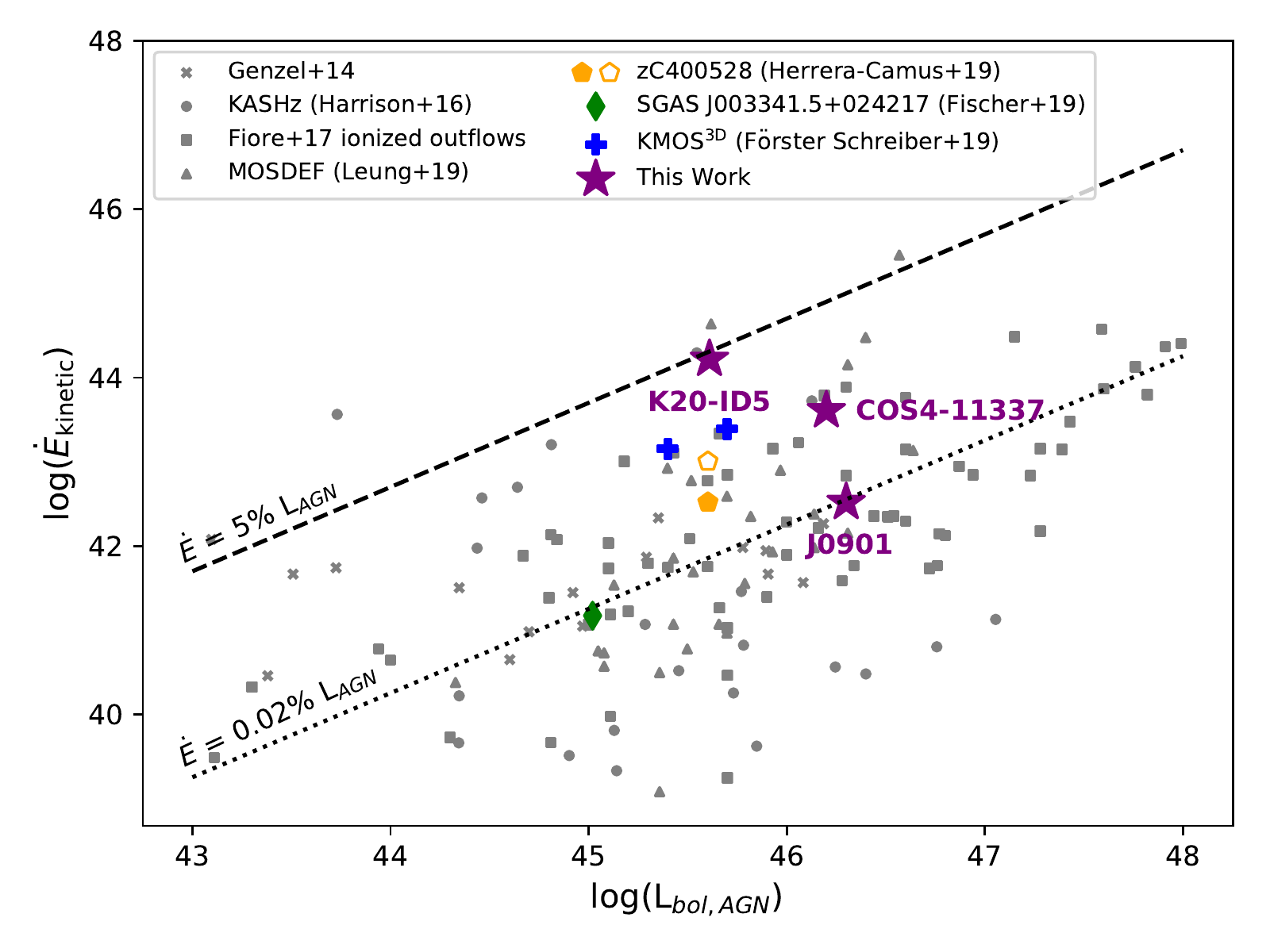}
\caption{Ionized outflow kinetic power as a function of AGN bolometric luminosity for K20-ID5, COS4-11337 and J0901, compared to a compilation of AGN-driven outflows at $z\sim$~1-3 from the literature (see Section \ref{subsec:multiphase_outflows} for details). The literature values have been scaled to an electron density of $n_e$~=~1000~cm$^{-3}$. The dotted and dashed lines show the scalings for ionized outflows with kinetic power equivalent to 0.02\% and 5\% of the AGN bolometric luminosity, respectively. The filled and open points for zC400528 indicate measurements made from ionized gas only and from the sum of the ionized and molecular gas components, respectively. \label{fig:kinetic_energy_scaling}}
\end{figure*}

Even when outflows can be observed in multiple phases, constructing an accurate budget of the mass and energy in the different phases is very challenging. Ionized gas outflow masses scale with the inverse of the electron density, which is a relatively poorly constrained quantity. Recent studies suggest that the luminosity-weighted density of ionized gas in AGN-driven outflows is $\sim$~1000~cm$^{-3}$ \citep[e.g.][]{Perna17, Kakkad18, Husemann19, NMFS19, Shimizu19}, but many studies in the literature adopt n$_e$~$\lesssim$~100~cm$^{-3}$, indicating potential discrepancies on the order of magnitude level. CO-based molecular gas outflow rates scale with the \mbox{CO-to-H$_2$} conversion factor ($\alpha_{\rm CO}$), for which the typically adopted values vary between 0.8 (the `ULIRG' value; e.g. \citealt{Cicone14}) and 4.3 (the Milky Way value; \citealt{Bolatto13}), and optically thin outflows with even lower conversion factors have been reported in two objects \citep{Dasyra16, Lutz20}.

Although the exact distribution of mass and energy between different outflow phases is poorly constrained, it is clear that the mass outflow rates and mass loading factors listed in Table \ref{table:outflow_properties} and the outflow coupling efficiencies listed in Table \ref{table:outflow_energy_momentum} are only lower limits. This must be taken into consideration when evaluating the potential impact of outflows on the evolution of their host galaxies. 

The \mbox{M$_{\rm BH}$-$\sigma$} \citep{Ferrarese01} and \mbox{M$_{\rm BH}$-M$_{\rm bulge}$} \citep{Magorrian98} relations provide indirect evidence to suggest that black holes co-evolve with their host galaxies. The gravitational energy released by accretion onto supermassive black holes greatly exceeds the binding energy of the bulge, and therefore AGN feedback is widely considered an important mechanism for shaping this relationship. Analytical theories predict that AGN driven winds should have \mbox{$\dot{E}_{\rm out} \sim$~0.05~L$_{\rm edd}$}, and that this relationship should naturally give rise to the \mbox{M$_{\rm BH}$-$\sigma$} relation as a locus of balance between the momentum injection rate from the AGN and the gravitational potential of the bulge. Black holes above the M-$\sigma$ relation are predicted to drive galaxy-scale energy conserving outflows which eject gas from the bulge and prevent further black hole growth \citep[e.g.][]{King03, Zubovas12, Lapi14}. Various studies based on numerical simulations have reported that a 5\% coupling efficiency is sufficient to drive strong outflows which halt star formation and black hole growth, and leave galaxies on the M-$\sigma$ relation \citep[e.g.][]{DiMatteo05}.

Figure \ref{fig:kinetic_energy_scaling} shows the relationship between the outflow kinetic power and the AGN bolometric luminosity for K20-ID5, COS4-11337 and J0901. For comparison, we also plot data for a literature compilation of AGN-driven ionized outflows at $z\sim$~1-3 \citep{Genzel14, Harrison16, Fiore17, Leung19, HerreraCamus19, Fischer19, NMFS19}. All the literature measurements have been scaled to n$_e$~=~1000~cm$^{-3}$ for consistency.

There is a clear correlation between the AGN bolometric luminosity and the outflow kinetic power. The average ratio of the outflow kinetic power to the AGN bolometric luminosity is 0.02\% (black dotted line). There is a large scatter around the average (primarily driven by variations in outflow velocity at a fixed AGN luminosity), but in the vast majority of cases the coupling efficiency is well below the 5\% level suggested by the models (black dashed line). K20-ID5 is one of the extreme cases falling close to the 5\% line, but the coupling factor for COS4-11337 is a factor of ten lower at 0.3\%. J0901 is another factor of ten lower at 0.02\%, but this is not surprising given the likely momentum conserving nature of the outflow. Even if we were to assume an electron density of 100~cm$^{-3}$, the average coupling factor for the full sample would be 0.2\% - still a factor of 25 too low. The coupling between the AGN ionizing radiation field and the ionized gas outflows does not appear to be efficient enough for the M-$\sigma$ relation to be the consequence of self-regulating black hole feedback. \citet{AnglesAlcazar13} showed that a torque-limited accretion model (in which the inflow rate onto the black hole accretion disk is driven by gravitational instabilities in the galaxy disk) naturally reproduces the M-$\sigma$ relation without the need for any coupling between the AGN accretion energy and gas on galaxy scales.

Accounting for the mass and energy in other phases of the outflows would result in higher $\dot{E}_{\rm out}$ values and may partially alleviate the discrepancy with first order expectations for self-regulated black hole growth. However, for zC400528, a normal AGN host galaxy at $z\sim$~2, the molecular and ionized phases have similar kinetic power \citep{HerreraCamus19}, and therefore the overall coupling efficiency does not change significantly depending on whether only the ionized phase (filled yellow pentagon in Figure \ref{fig:kinetic_energy_scaling}) or both the ionized and molecular phases (open yellow pentagon) are considered. Further multi-phase studies of outflows in individual galaxies as well as better constraints on uncertain parameters such as $n_e$ and $\alpha_{\rm CO}$ will be crucial for gaining further insights into the primary mode of black hole growth and the degree of coupling between the AGN accretion energy and gas in the host galaxy.

\subsection{Strong AGN-Driven Outflows in Compact Star Forming Galaxies}
K20-ID5 and COS4-11337 are particularly interesting systems because they provide insights into the role of AGN feedback in driving the evolution of compact star forming galaxies (cSFGs). cSFGs lie on or above the SFR main sequence but have sizes more similar to those of compact quiescent galaxies. There is growing evidence that the build up of large stellar mass surface densities is closely linked to the quenching of star formation \citep[e.g.][]{Martig09, Bluck14, Lang14}, and cSFGs may represent an intermediate population of galaxies that have already undergone morphological transformation but have not yet ceased forming stars \citep[e.g.][]{Barro13, Nelson14, Williams14}. Most cSFGs appear to have lower gas fractions and shorter depletion times than normal star forming galaxies, suggesting that they will indeed quench on relatively short timescales \citep[e.g.][]{Barro16b, Spilker16, Popping17, Tadaki17, Talia18}.

Compaction occurs when a large amount of gas is funneled towards the center of a galaxy - for example as a result of disk instabilities \citep[e.g.][]{Bournaud07, Dekel14, Brennan15} or galaxy-galaxy interactions \citep[e.g.][]{Hopkins10}. The presence of a large nuclear gas reservoir can trigger star formation and/or AGN activity, which can subsequently drive outflows. cSFGs exhibit a higher incidence of AGN activity than normal star forming galaxies at fixed stellar mass \citep{Kocevski17}, and are also expected to have a high incidence of AGN-driven outflows based on their large stellar masses and central stellar mass densities \citep[e.g.][]{NMFS19}. Therefore, it is important to consider the potential role of AGN feedback in quenching star formation in cSFGs. 

K20-ID5 and COS4-11337 are both classified as `Strong Outflows' by \citet{NMFS19} because an unusually large fraction of their \NII+\Ha\ emission is associated with their $\sim$1500~\kms\ AGN-driven outflows. This is exemplified in K20-ID5, for which our analysis suggests that almost all of the nuclear line emission is associated with the outflow. Strong Outflows are rare, occurring in $\sim$5\% of massive (log(M$_*$/M$_\odot$)~$\gtrsim$~10.7) galaxies, and accounting for $\sim$10\% of AGN-driven outflows. They have similar outflow velocities and global mass loading factors to normal AGN-driven outflows, but have $\sim$~2.5$\times$ higher mass outflow rates and are found in galaxies that are smaller and have higher SFRs and specific AGN luminosities (sometimes used as a proxy for Eddington ratio) compared to typical AGN host galaxies at the same redshift. Strong Outflows may therefore trace a `blowout' phase which is also associated with strong SF and black hole accretion activity.

The impact of these extreme outflow phases on SF in the host galaxy is unclear. The sub-dominant contribution of SF to the \Ha\ emission in the nuclear regions could indicate either that there is very little nuclear SF, or that the nuclear SF is heavily obscured, which would be expected if the gas mass surface densities in the central regions are high. The Strong Outflows on average have similar global mass loading factors to normal AGN-driven outflows, suggesting that the SF activity in the host galaxy is relatively unaffected by the extreme nuclear blowout, at least in the early stages. However, the outflows carry significant amounts of kinetic energy into the circumgalactic medium, which may help to maintain the presence of a hot halo and therefore impede replenishment of the molecular gas reservoir \citep[e.g.][]{Bower06, Croton06, Bower17, Pillepich18}.

The outflows from K20-ID5 and COS4-11337 decrease the already short molecular gas depletion times in these systems. K20-ID5 has a molecular gas mass of \mbox{log(M$_{H_2}$/M$_\odot$) = 11.0} (calculated by re-scaling the CO-based gas mass from \citealt{Popping17} to the metallicity-dependent $\alpha_{CO}$ from \citealt{Tacconi18}), corresponding to a SF depletion time of 280~Myr. If we assume that the molecular gas outflow rate is at least as large as the ionized gas outflow rate (see discussion in Section \ref{subsec:multiphase_outflows}), the overall depletion time (including the contribution of the outflow) is $\leq$~160~Myr, compared to an average depletion time of 520~Myr for galaxies at the same stellar mass, SFR and redshift \citep{Tacconi18}. There are no existing gas mass measurements for COS4-11337, but using the upper limit on the \mbox{CO(4-3)} flux (Figure \ref{fig:cos4_11363_co43}), we find a 3$\sigma$ upper limit on the gas mass of \mbox{log(M$_{H_2}$/M$_\odot$) $<$ 9.9} (assuming \mbox{CO(1-0)/CO(4-3) = 2.4}, and the metallicity-dependent $\alpha_{CO}$ from \citealt{Tacconi18}). The gas mass upper limit corresponds to a SF depletion time of \mbox{$<$~18~Myr}, and an overall molecular gas depletion time of $<$~16~Myr, compared to an average of 540~Myr for galaxies at the same stellar mass, SFR and redshift. The molecular gas depletion time for \mbox{COS4-11337} is very short, and may indicate that the AGN radiation field is heating some of the molecular gas and causing it to emit primarily in higher excitation CO transitions \citep[e.g.][]{Gallerani14, Mingozzi18, Rosario19}. 

Both K20-ID5 and \mbox{COS4-11337} could deplete their entire molecular gas reservoir within a couple of hundred Myr, and the kinetic energy injected into the halos by the AGN-driven outflows could suppress the accretion of fresh molecular gas, supporting the notion that these galaxies may be the direct progenitors of compact quiescent systems. However, both galaxies are currently located on the upper envelope of the star-forming main sequence, suggesting that the outflows have not yet had any significant impact on the SF activity in their host galaxies.

\section{Summary and Conclusions}
\label{sec:conc}
We have used deep SINFONI-AO data to characterize the AGN-driven outflows in three massive \mbox{(log(M$_*$/M$_\odot$)~$\sim$~11)} main sequence galaxies at $z\sim$~2.2 -- K20-ID5, COS4-11337 and J0901. These galaxies probe AGN feedback acting on nuclear, disk, and circumgalactic scales, and therefore provide important insights into the different mechanisms through which AGN-driven outflows can interact with their host galaxies and surrounding environment.

K20-ID5 has a luminous compact core and a fainter, regularly rotating extended disk. Our SINFONI-AO data reveal strong deviations from regular disk kinematics in the central 0.4'' (3.3~kpc), spatially coincident with elevated line widths and large \NIIHa\ ratios. We conclude that the majority of the line emission in the nuclear region traces the AGN-driven outflow, with a minor contribution from star formation. The outflow can be traced well beyond the effective radius of the galaxy, to a distance of $\sim$5~kpc, at an approximately constant velocity of $\sim$1400~\kms. 

COS4-11337 is a compact star forming galaxy in a close pair with COS4-11363, at a projected separation of only 5.4~kpc. COS4-11337 shows very strong and broad line emission, whereas \mbox{COS4-11363} has very little line emission and is likely to have experienced a rapid decrease in SFR in the last hundred Myr. We identified \mbox{CO(4-3)} emission at the location of \mbox{COS4-11363} in archival ALMA data, confirming that it lies close in velocity space to COS4-11337 \mbox{($\Delta v <$~150~kms)}. The SINFONI-AO spectrum of COS4-11363 reveals a very high \NIIHa\ ratio of 2.6, indicative of shock excitation. We showed that the $\sim$1500~\kms\ outflow driven by the AGN in \mbox{COS4-11337} is propagating towards \mbox{COS4-11363} and may therefore be responsible for shock heating the ISM in the companion galaxy. However, we cannot rule out a scenario where most or all of the shock excitation in COS4-11363 is due to tidal torques induced by the galaxy-galaxy interaction.

The outflows in K20-ID5 and COS4-11337 have small half-light radii ($\sim$1 kpc) but can be traced to large galactocentric distances ($\gtrsim$~5~kpc). Combined with previous results, this suggests that AGN-driven outflows have steep luminosity profiles, with luminous cores and faint extended tails, perhaps driven by a decrease in the surface brightness and/or density of the outflowing material as it propagates away from the galaxy nuclei.

K20-ID5 and COS4-11337 are unique objects because they are classified as compact star forming galaxies and show abnormally strong outflow signatures in their nuclear spectra. The lack of prominent galaxy emission in the nuclear regions may indicate that there is very little nuclear star formation or that the nuclear region is heavily obscured, the latter of which is plausible in the case of high nuclear gas mass surface densities. The galaxies are located on the upper envelope of the star forming main sequence, suggesting that the outflows do not have any significant impact on the instantaneous SF activity in their host galaxies. However, the outflows carry a large amount of kinetic energy which will be injected into the circumgalactic medium, and could contribute to the maintenance of a hot halo. The resulting suppression of cold gas accretion combined with the already short depletion times in these systems could perhaps lead to rapid exhaustion of the molecular gas reservoirs on timescales of a few hundred Myr. 

The outflow in J0901 has very different properties to the outflows in K20-ID5 and COS4-11337. J0901 is gravitationally lensed, providing us with a magnified view of the nuclear region. Despite the factor of two enhancement in spatial resolution, the nuclear outflow is barely resolved, and has a half light radius of 470~$\pm$~70~pc and a velocity of $\sim$~650~\kms.

The AGN in J0901 has a similar luminosity to the AGN in COS4-11337, and therefore the difference in outflow extent and velocity does not appear to be related to the \textit{current} AGN luminosity. We postulate that the J0901 outflow may be in an early stage of its evolution (i.e. it has not yet had sufficient time to break out of the nuclear region), or the conditions in the nuclear region may lead to inefficient coupling between the AGN radiation field and the gas in the host galaxy. In its current state, the outflow in J0901 is not able to transfer a significant amount of mass or energy out of the nuclear region. However, the dissipation of kinetic energy from the outflow could potentially increase the turbulence in the circumnuclear region enough to stabilise molecular gas against collapse, and therefore decrease the star formation efficiency in the center of the galaxy.

Finally, we investigated whether the efficiency of the coupling between the AGN radiation field and the ionized gas outflows is sufficiently strong for the M-$\sigma$ relation to be explained by self-regulating black hole feedback. We combined our measurements with a  compilation of ionized outflows at $z\sim$1-3 from the literature, and found an average coupling factor (\Edotout/L$_{\rm AGN}$) of 0.02\%. \mbox{K20-ID5} has one of the highest coupling factors in the sample at 4\%, J0901 lies at the average value of 0.02\%, and COS4-11337 has an intermediate coupling factor of 0.3\%. The low average coupling factor may lend support to alternative origins for the M-$\sigma$ relation such as torque-limited black hole accretion. However, we emphasise that there are many uncertainties in the calculation of the coupling factors. The outflow kinetic energy scales inversely with the electron density, which is a poorly constrained quantity. In this work we have only probed the ionized gas phase of the outflows, and it is unclear what fraction of the outflow kinetic power is carried in the molecular, neutral and hot phases. In the future it will be critical to gather a large sample of AGN-driven outflows with robust measurements of the outflow mass in multiple gas phases, in order to better determine what fraction of the mass and kinetic power is found in the ionized phase, and whether this varies as a function of AGN luminosity or other galaxy properties. 

\acknowledgements
We thank the referee for a constructive report which improved the clarity of this paper. Based in part on observations collected at the European Organisation for Astronomical Research in the Southern Hemisphere under ESO Programme IDs 074.A-9011, 092.A-0082, 092.A-0091, 093.A-0079, 093.A-0110, 094.A-0568, 095.A-0047, 097.B-0065, 0101.A-0022. Based in part on data obtained at the LBT. The LBT is an international collaboration among institutions in the United States, Italy and Germany. LBT Corporation partners are: The University of Arizona on behalf of the Arizona Board of Regents; Istituto Nazionale di Astrofisica, Italy; LBT Beteiligungsgesellschaft, Germany, representing the Max-Planck Society, The Leibniz  Institute for Astrophysics Potsdam, and Heidelberg University; The Ohio State University, and The Research Corporation, on behalf of The University of Notre Dame, University of Minnesota and University of Virginia. Based in part on observations made with the NASA/ESA Hubble Space Telescope, obtained from the data archive at the Space Telescope Science Institute. STScI is operated by the Association of Universities for Research in Astronomy, Inc. under NASA contract NAS 5-26555. This paper makes use of the following ALMA data: ADS/JAO.ALMA\#2015.1.00228.S, 2016.1.00726.S. ALMA is a partnership of ESO (representing its member states), NSF (USA) and NINS (Japan), together with NRC (Canada), MOST and ASIAA (Taiwan), and KASI (Republic of Korea), in cooperation with the Republic of Chile. The Joint ALMA Observatory is operated by ESO, AUI/NRAO and NAOJ. This research made use of \textsc{Astropy}, a community-developed core Python package for Astronomy \citep{Astropy13, Astropy18} and \textsc{Matplotlib} \citep{Hunter07}.

\appendix

\section{J0901: Lens Modelling, Source Plane Reconstruction and Physical Properties}

\subsection{Lens Modelling with \textsc{Lenstool}}\label{appendix:lens_modelling}
\begin{figure}
\centering
\includegraphics[width=0.5\hsize]{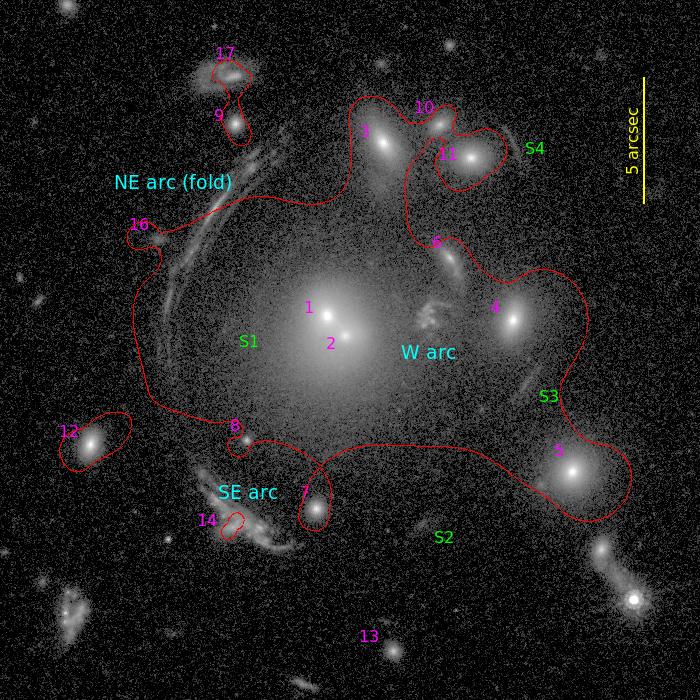}
\caption{HST F814W image of J0901, with overplotted information related to the lens modelling. N is on top. Cyan labels identify the three J0901 arcs. S1 to S4 (green labels) label the four images of the $z\approx 3.1$ `Sith' 
lensed background object. Foreground cluster objects used in the lens model are labelled in magenta. No. 14 is the perturber near the SE arc. The outer critical line for the J0901 redshift is overplotted in red.}
\label{fig:overview}
\end{figure}

J0901 at $z=2.2586$ \citep{Hainline09} is lensed by a poor cluster at $z\sim 0.3459$ \citep{Diehl09}. Three images are apparent in Figure~\ref{fig:overview}: the southeast (SE) arc that is the target of our high resolution SINFONI observations, a relatively undistorted western (W) arc, and the northeastern (NE) arc which is a fold arc that images part of the source twice, but misses the H-band nucleus \citep[see also][]{Tagore14,Sharon19}. To construct a lens model for interpretation of the SINFONI adaptive optics and HST data, we rely on archival HST F814W (rest frame UV) and F160W (rest frame optical) images, and use version 7.0 of the parametric gravitational lens modelling code \textsc{lenstool} \citep{Kneib96,Jullo07,Jullo09}. All HST data have been astrometrically registered for consistency with each other and with five stars from GAIA DR1 \citep{Gaia16}.

\begin{figure}
\centering
\includegraphics[width=0.35\hsize]{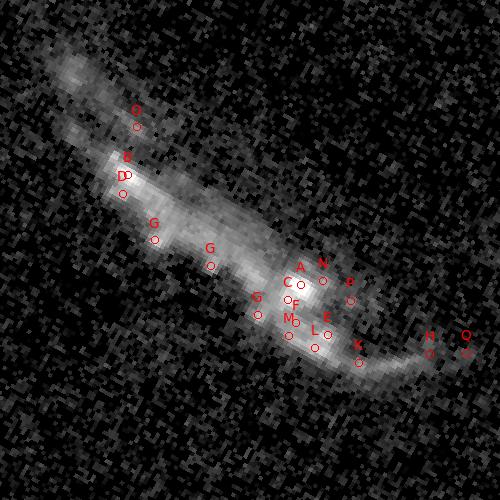}
\includegraphics[width=0.35\hsize]{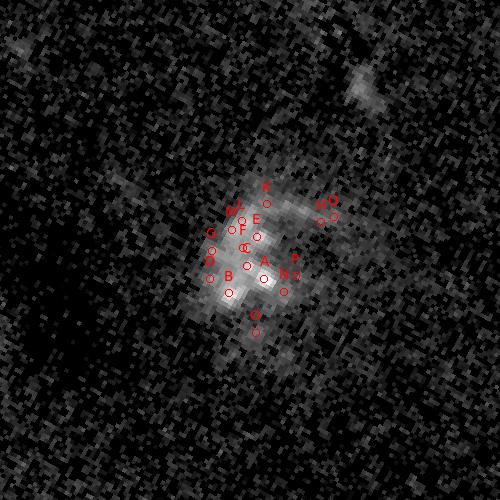}
\caption{Identification of I-band knots used to constrain the lens model, overplotted on foreground object subtracted F814W images of the SE arc (left) and W arc (right). Note that the H-band nucleus coincides with knot C rather than the I-band brighter knot A. Knot G is triply imaged in the SE arc, due to the nearby foreground perturber galaxy which is already subtracted for this image. Residual from foreground galaxy subtraction are visible towards the top of the right panel.}
\label{fig:input}
\end{figure}

The model is constrained by the locations of 15 I-band (rest UV) emission knots (Figure~\ref{fig:input}), with positions that have been manually measured in the W arc, in the SE arc (one of them imaged three times within the SE arc), and for five of them in the NE fold arc (all of these imaged twice). In addition, we use the four HST H-band images of a reddish second lensed object with unknown redshift, nicknamed `Sith' by \citet{Tagore14}. The redshift of this object is left free for the lens modelling and is estimated at $z\approx 3.1$ by the adopted \textsc{lenstool} model.

The lens model includes a general cluster potential for which we adopt an NFW profile \citep{Navarro97}. We fix the NFW concentration parameter at $c=6$, reasonable for a $M_{200}\approx 10^{14}$~M$_\odot$ cluster reported below \citep{Dutton14,Merten15,Umetsu16}, and considering some bias towards more concentrated halos for strong lensing clusters. The position, axial ratio, position angle and radius of the cluster potential are left free for the fit. We note that our data do not strongly test or constrain this adopted cluster radial profile shape, since all J0901 and Sith constraints are at similar cluster-centric radii. 

\begin{table*}
\begin{nscenter}
\caption{Best-fit lens model parameters.}
\begin{tabular}{lccccccc}
\hline
Halo     & RA    & DEC   & e   & PA  &$r_{core}$&$r_{cut}$&$\sigma_0$\\
         &\arcsec&\arcsec&     & (deg) & (kpc)       & (kpc)      & (\kms) \\ \hline
Cluster  &  -1.53&   1.48&\multicolumn{5}{c}{NFW e=0.32, PA=82.7, [c=6], $r_{sca
le} = 127$~kpc,  $M_{200}=7.3\times10^{13}$~M$_\odot$}\\
Galaxy 1 &[-0.79]& [2.13]& 0.22& 35.4&[0.25]    &211.5    &270.0\\
Galaxy 2 &[-1.51]& [1.33]& 0.22& 77.0&[0.25]    &221.5    &172.6\\
Perturber& [2.93]&[-6.25]& 0.01& 77.0&[0.25]    &384.6    & 85.9\\  
Galaxy scaling  &\ldots &\ldots &\ldots&\ldots&[0.15]  &  5.5    &235.9\\
\hline
\end{tabular}
\end{nscenter}
\tablecomments{Coordinates are in arcsecond relative to a fiducial RA~135.343500~deg DEC~18.241792~deg (ICRS). Values in square brackets are fixed to the input. Ellipticities are $e=(a^2-b^2)/(a^2+b^2)$. All Halos except the NFW `cluster' one are dPIE potentials. Positions angles are quoted degrees east of north (i.e., not in the \textsc{lenstool} convention). `Galaxy scaling' values refer to those galaxies where $r_{cut}$ and $\sigma_0$ are not individually fitted but scaled  $\propto L^{1/4}$. Values listed refer to a galaxy with H=17.3mag.}
\label{tab:lensmodel}
\end{table*}

Galfit \citep{Peng02,CYPeng10} was applied to the HST I-band and H-band images in order to derive parameters of the foreground cluster members. For the two galaxies near the cluster center, we adopt dPIE \citep{Eliasdottir07} profiles with fixed position and core radius but fit the axial ratio, position angle, velocity dispersion, and cut radius. Close proximity to the cluster center makes their individual parameters less constrained. For the perturber galaxy near the J0901 SE arc, we fix position and position angle to the I band observed values but fit axial ratio, velocity dispersion, and cut radius. For 13 additional foreground cluster galaxies, we follow the common procedure \citep[e.g.,][]{Limousin07} of fixing position, position angle, and axial ratio individually to the I band observed
values, but fitting $\propto L^{1/4}$ relations for velocity dispersion and cut radius, with L based on the H band magnitude. Finally, we use the crossing of the NE fold arc by the critical line and the arc orientation at that point as constraints. Table~\ref{tab:lensmodel} lists resulting parameters.

\begin{figure}
\centering
\includegraphics[width=0.6\hsize]{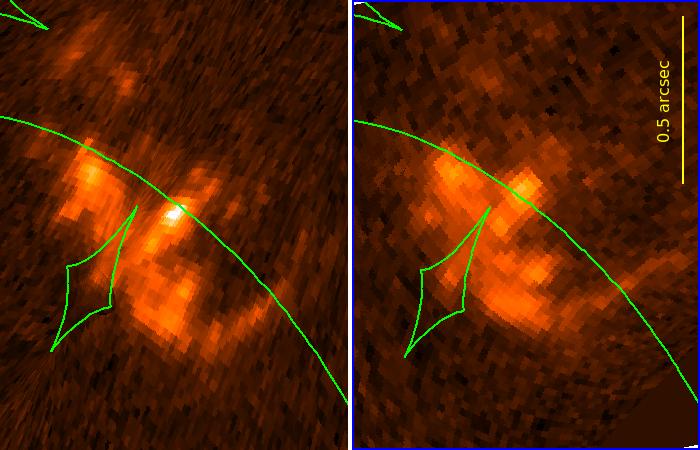}
\caption{I band source plane projections of the SE arc (left) and W arc (right). The source plane image for the W arc has been scaled $\times$0.955 and rotated by 11.6\degr\/. Outer caustics are overplotted in green, the small diamond-shaped caustic is related to the perturber near the SE arc and encloses knot G.}
\label{fig:i_source}
\end{figure}

The model fits the overall lens morphology. The RMS difference in the image plane between observed position of a knot and position predicted from the knot's location in another arc plus lens model is 0.45\arcsec\/. This reduces to 0.16\arcsec\ when comparing only SE and W arc, i.e. excluding the strongly distorted NE arc and Sith. The overall morphology and in particular the need to consider the perturber near the SE arc match the findings that \citet{Sharon19} obtained from a lens model that is based on 1.1\arcsec\ resolution IRAM CO data. Since our focus is on interpretation of higher spatial resolution SINFONI adaptive optics data, we adopt the model built from high resolution HST data. Figure~\ref{fig:i_source} provides a sanity check of the lens model, by comparing the SE and W I-band arcs, projected back to the source plane and applying a small scaling/rotation correction to the W reconstruction. Before applying the lens model to SINFONI data, we shifted the astrometry of the SINFONI cubes to obtain a K-band continuum position consistent with the HST F160W image.

\subsection{AGN Luminosity}\label{appendix:j0901_lagn}
We calculate the luminosity of the AGN in J0901 from the \OIII\ luminosity. It is not possible to directly compute the source plane \OIII\ luminosity because we only have long slit data. Instead, we assume that the \NII\ and \OIII\ emission have similar spatial distributions, and multiply the image plane \OIII\ luminosity by the ratio of the \NII\ flux in the source plane nuclear spectrum to the \NII\ flux in the image plane long slit spectrum. We correct the \OIII\ luminosity for extinction using the Balmer decrement (described in Section \ref{subsec:j0901_energetics}), and convert the \OIII\ luminosity to the AGN bolometric luminosity using a bolometric correction factor of 600 \citep{Netzer09}, yielding an AGN luminosity of \mbox{log(L$_{\rm AGN}$) = 46.3}.

\subsection{Stellar Mass}\label{appendix:j0901_mstar}
\citet{Saintonge13} estimated the stellar mass of J0901 by making use of empirical correlations between the 3.6$\mu$m and 4.5$\mu$m luminosities and the stellar mass. The observed luminosities were corrected for lensing magnification using a single wavelength-independent correction factor. We re-calculate the stellar mass and $A_V$ for J0901 by utilising archival HST imaging in the F475W ($\lambda_{rest}$~=~1481\AA), F814W ($\lambda_{rest}$~=~2500\AA) and F160W ($\lambda_{rest}$~=~4959\AA) bands (Program ID 11602, PI: S. Allam). The imaging has a spatial resolution of 0.1-0.15'', which is sufficient to apply the source plane reconstruction and calculate the magnification factor for each band individually. We calculate the source plane luminosities in these three bands and fit the SED using the \textsc{FAST} code \citep{Kriek09}. We consider solar metallicity models from the \citet{BC03} library, with dust extinction following the \citet{Calzetti00} curve and $A_V$~=~0-3. We assume an exponentially declining star formation history ($\tau$ model), allowing for ages between 50~Myr and 2.86~Gyr (the age of the universe at $z$~=~2.259), and log($\tau$/yr)~=~8.5-10.0. The best fit model gives \mbox{log(M$_*$/M$_\odot$)~=~11.2} and \mbox{$A_V$~=~1.2}. Our stellar mass estimate is close to the value that is obtained by scaling the \citet{Saintonge13} value to our \Ha\ magnification factor (\mbox{log(M$_*$/M$_\odot$)~=~11.03~$\pm$~0.14}).

\subsection{SFR}\label{appendix:j0901_sfr}
We calculate the SFR of J0901 from the 160$\mu$m flux presented in \citet{Saintonge13}. We scale the total 160$\mu$m flux by a factor of 0.4 to isolate the flux originating from the SE arc \citep{Saintonge13}. We cannot measure the magnification factor in the FIR, because the spatial resolution of the imaging is insufficient to perform the source plane reconstruction. Instead, we adopt the \Ha\ magnification factor ($\mu$~=~9.9), which provides a good first order approximation of the lensing correction. We convert the 160$\mu$m luminosity to L$_{\rm IR}$ using the SED template from \citet{Wuyts08}, and convert the L$_{\rm IR}$ to a SFR using \mbox{SFR = 1.09~$\times$~10$^{-10}$~L$_{\rm IR}$/L$_\odot$} \citep{Wuyts11}. The resulting SFR is 200~M$_\odot$~yr$^{-1}$.

\bibliography{../bibliography/mybib}

\end{document}